\PassOptionsToPackage{dvipsnames}{xcolor}	
\documentclass[12pt,letter]{article}

\usepackage[T1]{fontenc}
\usepackage{lmodern}
\usepackage{setspace}
\usepackage{etoolbox}
\makeatletter
\patchcmd{\appendix}{\@Alph}{\@Roman}{}{}
\makeatother

\usepackage{amsmath}
\usepackage{amssymb}
\usepackage{amsthm}
\usepackage{mathtools}
\usepackage{mathrsfs}
\usepackage{breqn}
\usepackage{bigints}
\usepackage{bbm}

\usepackage[margin=1 in]{geometry}
\usepackage{enumerate}
\usepackage{enumitem} 
\usepackage{xcolor}
\usepackage[hidelinks]{hyperref}
\usepackage{float}
\usepackage{indentfirst}
\usepackage{caption}
\usepackage{fancyhdr}
\usepackage{tcolorbox}
\usepackage{multirow,array}
\usepackage{bigints}

\usepackage{tikz}
\usetikzlibrary{decorations.pathreplacing}
\usepackage{mathtools}
\usetikzlibrary{positioning}
\usepackage{graphicx}
\tikzset{
    mysquare/.style={x=12cm, y=6cm, >=stealth}
}

\renewcommand{\epsilon}{\varepsilon}
\newcommand{\E}[2][]{\mathbb{E}_{#1}\left[#2\right]}

\newtheorem{theorem}{Theorem}
\newtheorem{lemma}{Lemma}
\newtheorem{proposition}{Proposition}

\newtheorem*{statement*}{Result}

\theoremstyle{definition}

\DeclareMathOperator*{\argmax}{\arg\!\max}

\DeclareTextFontCommand{\emph}{\slshape}

\usepackage{natbib}
\nocite{*}

\title{Multidimensional Sequential Screening}

\author{Eric Gao\thanks{Department of Economics, Massachusetts Institute of Technology. ericgao@mit.edu.} 
\footnote{I would like to thank Josephine Auer, Ian Ball, Drew Fudenberg, Andrew Koh, Daniel Luo, Parag Pathak, Eric Tang, Alex Wolitzky, Bryant Xia, and participants at the 3B Theory Conference, MIT Theory Lunch, and NEEPC for helpful feedback and discussions. \url{Refine.ink} was used to proofread the paper for consistency and clarity.}}

\date{\today}

\begin{document}

\maketitle
	
\begin{abstract}
    I study multidimensional sequential screening. A monopolist contracts with a buyer who privately observes information about the distribution of their eventual valuations for multiple goods. After initial private information is reported and the contract is signed, the buyer learns and reports realized valuations. In these settings, the monopolist frontloads surplus extraction: Any information rents given to the buyer to elicit their true valuations can be extracted in expectation before those valuations are drawn, transforming the multidimensional screening problem by distorting buyer information rents compared to static screening. If the buyer’s distributions over valuations are commonly FOSD ordered, regular for each good, and satisfy \textit{invariant dependencies} (valuations can be dependent across goods, but how valuations are coupled cannot vary), the optimal mechanism coincides with independently offering the optimal sequential screening mechanism for each good separately. This rationalizes membership payments followed by separate sales schemes commonly used in practice.
\end{abstract}

	
\newpage
\onehalfspacing

\pagenumbering{arabic}

\section{Introduction}

Firms often contract with consumers across multiple products and across multiple time periods. How do these distinct sources of multidimensionality---across space and time---interact and influence the design of optimal mechanisms? One strategy often used in practice is options-style pricing, where consumers are given the option to pay some upfront fee to receive lower prices later on. This scheme is seen in multiple industries. In retail, consider  Costco's ``Gold Star'' versus ``Executive'' memberships, Sam's Club ``Club'' versus ``Plus'' memberships, or Amazon's ``standard'' versus ``Prime'' accounts. In each case, the higher tier comes at a higher upfront payment but lower eventual costs in the form of cash-back on purchases, monthly credits, or free shipping. In air transportation (the classic application of sequential screening as introduced in \cite{CourtyLi2000}), many airlines offer the option to pay a higher upfront price to make flights refundable. In cloud computing, Google Cloud offers lower rates when committing to one or three years of usage compared to spot prices.

Another striking feature of such pricing schemes is that eventual spot prices are independent across goods. The price of eggs does not depend on milk purchases; the price of flights to Athens does not depend on planned trips to Barcelona; the price of M2 memory-optimized machine usage does not depend on usage of C2 standard machines. This stands in sharp contrast to the (static) multidimensional screening literature, which finds bundling (often in complicated and intractable ways) to be optimal. Why is it, then, that products are not bundled in practice?

To answer these questions, I study a model of dynamic multidimensional screening where a monopolist contracts with a buyer endowed with private information about their distribution over eventual (multidimensional) valuations. In period one, the contract is written while the buyer only knows their distribution over eventual valuations. In period two, actual valuations are realized, goods are allocated, and transfers are made. The model generalizes the canonical sequential screening setting of \cite{CourtyLi2000} to multiple goods, allowing us to study the interaction between dynamic and multidimensional screening. In this setting, the optimal mechanism is to offer the buyer the option to pay some amount upfront to secure lower eventual posted prices to purchase each good on its own (Theorem \ref{no_bundlding}).\footnote{This is one way to implement the optimal allocation rule; in general payments can be shifted between periods one and two so an equivalent implementation could have all transfers occur in period two. In general, individuals who think their valuations will be high pay more fixed costs and less variable costs.}

To build intuition behind the optimality of separate sales, consider why bundling arises in (static) multidimensional screening. In general, bundling goods helps the monopolist manage incentive constraints and reduce information rents given to the buyer, thereby capturing a larger share of total surplus generated from trade. However, endowing the monopolist with the ability to screen through time provides them with an alternative tool to reduce the effective information rents given to the buyer to elicit their eventual valuations: Any such information rents can be extracted \textit{in expectation} before the buyer's valuations are realized. Thus the only information rents that the buyer can actually capture are those for their pre-contractual private information (formalized in Lemma \ref{formulation}). This frontloading of surplus extraction is not only \textit{feasible} per the previous discussion, but also \textit{optimal}. The principal needs to provide less information rents to the buyer when the buyer simply has less private information: Joint deviations have less bite as period one deviations can only be ex-ante coordinated with anticipated future deviations.\footnote{Frontloading surplus extraction is somewhat analogous to backloading the provision of information rents, similar to how transfers are often backloaded in dynamic moral hazard problems (see \cite{Ray2002}, \cite{Liu2025DynamicRewardDesign}, or \cite{luo2025payingpersuading}; \cite{Ball2023DynamicInformation} considers backloading information revelation). }

Three assumptions on the underlying environment are needed for a clean solution. First, the buyer's initial private information can be ordered with, and thus parameterized by, a single dimension, with higher values corresponding to a higher (in the first-order stochastic dominance sense) distribution of eventual valuations for each good. While this does collapse the initial problem down to a one-dimensional screening problem, the space of outcomes is the set of all multidimensional screening mechanisms mapping realized valuations into allocations and transfers. As such, without the decomposition developed in this paper, the problem remains multidimensional. Second, the joint distribution of initial information and eventual valuations must satisfy a sequential analogue of regularity for each good, commonly assumed in the literature. Finally, while final valuations may be coupled or correlated in any manner, this dependency structure cannot vary as the buyer's initial private information varies. This new assumption, which I call \textit{invariant dependencies}, arises only in the multidimensional case (the correlational structure is trivially invariant when there is only one good). When these conditions hold, the optimal mechanism takes a simple form. The optimal mechanism for each good on its own is to charge the buyer some upfront price to secure some eventual strike price, where buyers with higher initial private information being charged more upfront but less at the time of purchase. The optimal mechanism overall is to aggregate the optimal mechanism for each good: Charge the buyer the sum of each good's upfront price to secure the same eventual strike prices for each good. 

How does invariant dependencies arise in practice? For Costco and other retailers, the buyer's initial private information may be how many meals each week they plan on cooking (or some other proxy of this such as income or hours spent working). Conditional on this, different individuals have similar dietary needs (roughly the same breakdown of produce versus protein versus carbohydrates). Individuals may also have different cravings but once again, such shocks are plausibly independent of the buyer's initial private information. In air transportation, the buyer's initial private information may be their expected value of the trip while their final valuation depends on whether or not they are sick the day of travel. In this case, whether or not someone gets ill at some particular point in time is independent of how much they value going on vacation. Finally, consider an artificial intelligence startup with initial private information about the scale of their startup (i.e. funds from venture capital firms, whether services are offered to enterprises or individuals). When it comes time to train their models, the particular machine learning architecture depends on how the state-of-the-art has evolved between conceptualization, funding, and training. How the technological frontier changes is independent of any particular startup's initial conditions but does influence how many of each type of machine the startup needs to train their models.\footnote{For example, Google recently announced \href{https://research.google/blog/turboquant-redefining-ai-efficiency-with-extreme-compression/}{TurboQuant}, a set of algorithms that allows for significant compression of AI models, thereby reducing memory needed.} As such, invariant dependencies is plausibly satisfied in many of the motivating examples, consistent with and rationalizing separate sales. 

Thus, the central contribution of this paper can be viewed as identifying a simple and economically interpretable condition that rationalizes separate sales in multidimensional screening. While there has been work rationalizing certain bundling schemes, less progress has been made explaining the prevalence of screening goods independently. Furthermore, the rationalization provided in this paper does not require appealing to behavioral simplicity, robustness, or approximate optimality. Beyond this economic insight, I provide new tools and insights to solve multidimensional sequential screening problems, departing from the standard approaches used to solve solely sequential or solely multidimensional problems.

The remainder of the paper proceeds as follows. Section \ref{example} works through a simple example and compares pure bundling with the optimal mechanism. Section \ref{related} discusses related literature while Section \ref{model} sets up the model. Section \ref{decoupling} solves for the optimal mechanism and discusses how each assumption comes into play. Section \ref{relax} considers relaxations of the three sufficient assumptions. In particular, the optimal mechanism under invariant dependencies is still implementable and gives the principal the same payoff under any alternative correlational structure (that may or may not satisfy invariant dependencies). This implies that a principal need not be concerned about the correlational structure between goods, providing a resolution to the puzzle that the optimal multidimensional screening mechanism is often detail-dependent. Section \ref{extensions} compares the model in this paper to alternative models of multidimensional sequential screening to illustrate the strength of frontloading surplus extraction: The problem can be relaxed in several ways favoring the monopolist yet the firm attains as much revenue in the current model as any of the relaxations. Finally, Section \ref{discussion} concludes.

\subsection{Illustrative Example}\label{example}

Suppose a buyer is considering purchasing goods $A$ and $B$. Their utility from $A$ is $v + \epsilon$ and their utility from $B$ is $v - \epsilon$ where $v$ is some initial private information (their demand for some sort of good) and $\epsilon$ is some later shock (idiosyncratic tastes in favor of one of the two goods). Initially, only $v$ is (privately) known to the buyer when the contract is written; $\epsilon$ is (privately) revealed later on. Suppose $v$ is uniform over $[0, 1/2]$ and $\epsilon$ is uniform over $[-1/2, 1/2]$. The seller writes a direct revelation mechanism contracting on both $v$ and $\epsilon$.

Since the value of $A$ and $B$ are perfectly negatively correlated, the value of the bundle will always be $2v$ regardless of $\epsilon$. By only selling the grand bundle, the seller does not need to give the buyer any information rents to elicit $\epsilon$ and only needs to incentivize truthful elicitation of $v$. Under this regime, the problem turns into a one-dimensional screening problem selling to a buyer whose valuation of the bundle is uniform over $[0, 1]$. As such, the optimal mechanism is to post a price of $1/2$ for the grand bundle. Producer surplus is $1/4$, consumer surplus is $1/8$, and total surplus is $3/8$. Bundling obviates the need to pay information rents to elicit $\epsilon$ but comes at a cost of selling a good the consumer does not want. Can the seller do better? 

Instead of bundling, the seller can exploit the dynamic nature of the buyer's private information to reduce buyer information rents. Suppose the seller no longer bundled the two goods so information rents for $\epsilon$ still had to be paid. The seller knows the eventual information rents they need to give the buyer to elicit $\epsilon$. As such, the seller can instead extract \textit{expected} information rents \textit{before} the buyer knows what $\epsilon$ is. By doing this, the seller no longer needs to bundle; this avoids the issue of selling the buyer goods they do not want. 

Concretely, consider the following mechanism (which is indeed optimal via Theorem \ref{no_bundlding}). The buyer reports $v$ and pays $2v^2$ upfront. Then, $\epsilon$ is drawn and the seller posts a price of $1/2-v$ for each good independently (so the buyer can purchase neither good and pay zero, either one of the goods and pay $1/2-v$, or both goods and pay $1-2v$). Incentive compatibility for eliciting final valuations $v \pm \epsilon$ is immediate since the second-stage mechanism is a posted price mechanism, which is incentive-compatible. Thus, it is sufficient to verify incentive-compatibility to truthfully report $v$. We do so by computing the buyer's interim utility from potential misreports and use their first-order condition to verify truth-telling.

Suppose a buyer of type $v$ reports $\hat{v}$. They pay $2\hat{v}^2$ upfront and face a strike price of $1/2-\hat{v}$. The buyer thus purchases the first good if 
$$v + \epsilon \geq \frac{1}{2}-\hat{v} \implies \epsilon \geq \frac{1}{2}-\hat{v} - v$$
so their expected surplus from the first good is 
$$\int_{\frac{1}{2}-\hat{v} - v}^{\frac{1}{2}} (v + \epsilon - (1/2 - \hat{v})) d\epsilon = \left[ (v + \hat{v} - 1/2)\epsilon + \frac{\epsilon^2}{2} \right]_{\frac{1}{2}-\hat{v} - v}^{1/2} = \frac{1}{2}(v + \hat{v})^2.$$
By symmetry, this is also their expected surplus from the second good; as such the interim utility of a buyer with type $v$ reporting $\hat{v}$ is
$$\underbrace{\frac{1}{2}(v + \hat{v})^2}_{\text{Surplus from Good 1}} + \underbrace{\frac{1}{2}(v + \hat{v})^2}_{\text{Surplus from Good 2}} - \underbrace{2\hat{v}^2}_{\text{Upfront Payment}} = (v + \hat{v})^2 - 2\hat{v}^2.$$
The first order condition with respect to $\hat{v}$ is 
$$2(v+\hat{v}) - 4 \hat{v} = 0 \implies \hat{v} = v$$
and the second order condition is satisfied as $-2 < 0$. Thus incentive compatibility is satisfied. 

What are producer and consumer surplus from this mechanism? The seller gets $2v^2$ upfront and sells at each good at price of $1/2-v$ with probability $2v$ to a buyer of type $v$ and makes interim profits of 
$$2v^2 + 2(1/2-v)(2v) = 2v - 2v^2$$ 
to a type $v$ buyer. Thus, integrating over $v$ gives expected producer surplus to be 
$$\int_0^{1/2} (2v - 2v^2) 2dv = \left[2v^{2} - \frac{4v^{3}}{3}\right]_0^{1/2} =1/3$$
which is higher than the surplus of $1/4$ from the pure bundling mechanism. Interim surplus of a type $v$ buyer is $4v^2 - 2v^2 = 2v^2$ (by plugging in $\hat{v} = v$ into the expression for the surplus a buyer of type $v$ gets when reporting $\hat{v}$). Thus, expected consumer surplus is 
$$\int_0^{1/2} (2v^2) 2dv = 1/6$$
which is higher than the surplus of $1/8$ from the pure bundling mechanism. As such, this alternative mechanism increases both producer and consumer surplus; total surplus increases from $1/4+1/8 = 3/8$ to $1/3+1/6 = 1/2$.

\section{Related Literature} \label{related}

I relate to several strands of literature starting with the classic (separate) literatures on intertemporal screening and multidimensional screening. I also relate to recent work investigating the interaction between multidimensional screening and buyer information.

The sequential screening literature, started by \cite{CourtyLi2000}, considers a monopolist selling a single good to a buyer whose information arrives over time. \cite{EsoSzentes2017} generalizes the model to include buyer hidden actions and arbitrary time horizons and shows that the principal does not need to pay rents for information that arrives after the buyer enters the contract. For any implementable allocation rule, the cost of incentivizing the agent to reveal their entire sequence of private information is the same as the cost of incentivizing the agent to just reveal their initial private information. As such, downstream information is essentially observed ``for free'' as long as the desired mechanism is implementable. I apply this insight to the case of two periods but arbitrarily many goods in the second period and study the interaction of screening across time and space. The multidimensional problem is more difficult than the multiple-period problem since the buyer in my problem can coordinate deviations across different goods whereas the buyer in \cite{EsoSzentes2017} cannot base current deviations on future information. Furthermore, the irrelevance result of \cite{EsoSzentes2017} only holds among implementable allocation rules. When working in a multidimensional setting, the set of implementable allocation rules itself becomes more difficult to work with due to cyclic monotonicity constraints. As such, optimality of separate sales in period two is not a direct consequence given the insights of \cite{EsoSzentes2017}. Furthermore, absent invariant dependencies, the optimal mechanism may still involve nonlinear bundling, further emphasizing that multidimensionality has bite. As such, the insights developed in this paper are needed to make the multidimensional step tractable. See \cite{EsoSzentes2007}, \cite{segal2014}, \cite{KrahmerStrausz2017}, \cite{KrasikovLamba2022}, and \cite{li_2024_stochastic} for additional work on sequential screening. 

Next is the literature on multidimensional bundling, starting with \cite{Stigler1963}, \cite{AdamsYellen1976}, \cite{McAfeeMcMillanWhinston1989}, \cite{Armstrong1996}, and \cite{RochetChone1998}. Even when values are additive, the optimal mechanism features non-linear bundling and is generally intractable to find analytically (\cite{McAfeeMcMillan1988}, \cite{ManelliVincent2006}, \cite{Pavlov2011}, \cite{DaskalakisDeckelbaumTzamos2017}, \cite{RochetStole2003}, \cite{HartReny2015}, \cite{HartReny2019}, \cite{MenicucciHurkensJeon2015}). Another approach has been to \textit{start} with a particular bundling strategy and then find conditions on the underlying problem for which that strategy is optimal (\cite{HaghpanahHartline2021}, \cite{yang_2025_nested}). This paper is particularly connected to \cite{yang_2025_nested}, which also sets up the multidimensional screening problem with consumer heterogeneity ordered in a single dimension. \cite{yang_2025_nested} allows for non-additive valuations across goods but does not allow for sequential information arrival (thus also shrinking the set of potential mechanisms), which this paper does encompass.\footnote{After the monopolist collapses the multidimensional screening problem into a one-dimensional one via frontloading surplus extraction, revenues from selling incremental bundles given a fixed realization of final valuations are (1) single-crossing and (2) independent of other goods. As such, the optimal mechanism should exhibit nested bundling. In this paper, the optimal allocation rule still \textit{induces} nested bundling. Conditional on having the same realized final valuations, buyers that thought their distribution of valuations were higher pay more upfront and face lower eventual prices and thus choose to buy larger bundles at the prices they face.} There has also been a literature analyzing how well simple mechanisms (separate sales or selling the grand bundle) approximates the optimal mechanism (\cite{cai2012optimal}, \cite{li2013revenue}, \cite{yao2014reduction}, \cite{babaioff2014simple}, \cite{hart2017approximate}, \cite{cai2017simple}, \cite{cai2018duality}), but the bounds are multiplicative, the gap is large (and grows with the number of goods), and require goods to be identically and independently distributed.

Separate sales emerges in a multidimensional screening problem when robustness is a concern, developed in \cite{carroll2017} (and generalized in \cite{CheZhong2025}). In \cite{carroll2017}, the monopolist sells multiple goods and knows the buyer's marginal distribution of values for each good but would like to maximize the worst-case guarantee against an adversarially chosen correlational structure. In this setting, selling each good independently is optimal as the monopolist cannot exploit interdependencies between goods due to the nature of the robustness exercise: Bundled sales perform worse at some potential joint distribution. In this paper, instead of being \textit{unable} to exploit interdependencies, the monopolist does not \textit{need} to exploit interdependencies, as frontloading surplus extraction is simply more effective.

Finally, recent papers have combining buyer learning (over time) and bundling.\footnote{\cite{frick_2025_multidimensional} considers the seller learning instead. In their model, a multi-good monopolist observes informative signals (the seller perfectly knows the buyer's valuations as the number of signals goes to infinity) about a buyer's valuations. Pure bundling converges (as the number of signals goes to infinity) to full surplus extraction just as fast as the optimal mechanism whereas independent sales converges strictly slower. This paper, in contrast, endogenizes monopolist learning: Instead of exogenously observing informative signals, the buyer disclosing their initial information must be incentivized and is part of the contract.}  \cite{lu_2025_intertemporal} considers a monopolist selling two distinct goods, one in each of two distinct time periods. The buyer's initial information influences both their valuation of the first good and the distribution of values of the second good. If higher first-good valuations lead to a first-order stochastic higher distribution of second-good valuations then the optimal mechanism consists of the optimal static screening mechanism for the first good and the optimal sequential screening mechanism for the second good. However, if higher first-good valuations lead to a first-order stochastic \textit{lower} distribution of second-good valuations then the type for which individual rationality binds is endogenous and the optimal mechanism once again features bundling. Working with a single good in each period allows \cite{lu_2025_intertemporal} to abstract away from the cyclic monotonicity constraints that arise in multidimensional screening; this paper does not have any sales in the first period but an arbitrary number of goods in the second period. Individual rationality constraints being endogenous is why the distribution of valuations for each good must be ordered by first-order stochastic dominance \textit{in the same way} in my model: Without this assumption, individual rationality constraints become endogenous and different goods may be used to provide participation incentives, thus negating dynamic decoupling and making the problem much more difficult. Lastly, \cite{pernoud_2025_bundling} considers a simultaneous-move game where buyers choose to learn a one-dimensional signal about a multi-dimensional product while a monopolist chooses a selling mechanism. In equilibrium, buyers choose vertical learning (a signal that is positively correlated with each of the product's dimensions) while the seller's mechanism features nested bundling. In their model, the buyer's final purchasing decision is solely based on their signal while their true valuations are not part of the contracting problem. As such, frontloading surplus extraction is a key feature of my model not present in \cite{frick_2025_multidimensional} or \cite{pernoud_2025_bundling}.  

\section{Model}\label{model}

The model is a straightforward extension of \cite{CourtyLi2000} to a multidimensional setting. A monopolist contracts with a single buyer. The monopolist sells $n$ goods. Before contracting, the buyer observes $\gamma \in [\underline{\gamma}, \overline{\gamma}] =\Gamma$ drawn according to $G(\gamma)$. After contracting, the buyer draws $\theta \in \Theta = \prod_j [\underline{\theta}^j, \overline{\theta}^j] \subseteq \mathbb{R}^n$ according to $F(\theta|\gamma)$. I assume $F, G$ have strictly positive densities $f, g$ and $f$ is continuously differentiable with respect to $\gamma$ and $\theta$. Notationally, let superscripts denote marginals and subscripts denote partial derivatives, so $F^j_\gamma$ would denote the partial derivative of the marginal of $F$ in the $j$th dimension with respect to $\gamma$. 

The monopolist chooses $q: \Gamma \times \Theta \to [0, 1]^n, t_1: \Gamma \to \mathbb{R}, t_2: \Gamma \times \Theta \to \mathbb{R}$. The buyer chooses whether to participate in the mechanism and what $\gamma'$ to report if they do participate after observing $\gamma$ (but before observing $\theta$). Then, the buyer observes $\theta$ and decides what $\theta'$ to report. The monopolist's payoff is $t_1(\gamma') + t_2(\gamma', \theta')$. The buyer's payoff from receiving a (probabilistic) vector of goods $q$ while paying $t_1, t_2$ is 
$$q \cdot \theta - t_1 - t_2$$
where $\cdot$ is the dot product, so the buyer's value is additive across goods, there is no discounting across time, and $\gamma$ is payoff-irrelevant (except for determining the distribution of valuations). By the Revelation Principle, the monopolist's problem is:
\begin{align*}
    \max_{t_1, t_2, q} & \: \E[\gamma, \theta]{t_1(\gamma) + t_2(\gamma, \theta)} \\
    \text{s.t.} & \: \gamma \in \argmax_{\gamma'} \left\{\E[\theta|\gamma]{\max_{\theta'} q(\gamma', \theta') \cdot \theta - t_1(\gamma') - t_2(\gamma', \theta') }\right\} \text{ for all } \gamma \in \Gamma  ; & IC1 \\
    & \: \theta \in \argmax_{\theta'} q(\gamma, \theta') \cdot \theta - t_2(\gamma, \theta')  \text{ for all } \theta \in \Theta \text{ and } \gamma \in \Gamma; & IC2 \\
    & \: \E[\theta|\gamma]{q(\gamma, \theta) \cdot \theta - t_1(\gamma) - t_2(\gamma, \theta)} \geq 0 \text{ for all } \gamma \in \Gamma. & IR
\end{align*}
The constraints are as follows. First, $IC1$ pins down the buyer having no incentive to misreport $\gamma$ for any possible continuation misreport of $\theta$ after their new information arrives. However, their value of $\gamma$ still pins down the subsequent distribution of $\theta$. Next, $IC2$. After $\gamma$ is (truthfully) revealed, the buyer has no incentive to misreport $\theta$. Finally, $IR$ is an interim participation constraint: The buyer's choice of whether or not to participate in the contract can depend on their observed value of $\gamma$.

\subsection{Sufficient Assumptions}

The following three conditions on the underlying environment are sufficient for multidimensional sequential screening problems to have a simple solution of offering the optimal sequential screening mechanism independently in each dimension (Theorem \ref{no_bundlding}):
\begin{enumerate}
    \item[A1.] Commonly-Ordered: For all goods $j$ and $\gamma, \theta^j$, 
    $$F^j_\gamma(\theta^j|\gamma) \leq 0.$$
    That is, higher values of $\gamma$ lead to higher distributions of $\theta^j$ in the first-order stochastic dominance order for all goods $j$.
    \item[A2.] Sequential Regularity: For each good $j$, 
    $$\phi^j(\gamma, \theta^j) = \theta^j + \frac{F_\gamma^j(\theta^j | \gamma)}{f^j(\theta^j|\gamma)}\frac{1-G(\gamma)}{g(\gamma)}$$
    is weakly increasing in $\gamma$ and $\theta^j$.\footnote{Virtual values increasing in $\theta^j$ can be replaced with the weaker assumption of single-crossing instead.} That is, the sequential analogue of a buyer's ``virtual value'' satisfies regularity. This is standard in the sequential screening literature (for example, see \cite{EsoSzentes2007} or \cite{lu_2025_intertemporal}).
    \item[A3.] Invariant Dependencies: The distribution of percentiles 
    $$(F^1(\theta^1|\gamma), F^2(\theta^2|\gamma), ..., F^n(\theta^n|\gamma))$$
    is independent of $\gamma$. That is, valuations for each good can be coupled in arbitrary ways, but the coupling structure cannot vary as the buyer's initial private information varies. 
\end{enumerate}

None of the three assumptions are required to show that the buyer only needs to be provided information rents to elicit their value of $\gamma$. Assumption A3 then establishes that the monopolist's objective can be written to be additively separable in different goods. If, additionally, assumptions A1 and A2 hold, then the solution to the relaxed problem is implementable so global incentive constraints are not binding, obviating the previous concerns.

What do these assumptions imply? Using the orthogonalization approach of \cite{EsoSzentes2007} or \cite{EsoSzentes2017} makes the above three assumptions more easily interpretable. By Sklar's Theorem, there exists a unique distribution $C: [0, 1]^n \times \Gamma \to [0, 1]$ such that
$$F(\theta|\gamma) = C(F^1(\theta^1|\gamma),..., F^n(\theta^n|\gamma)|\gamma)$$
and letting $c$ be the density of $C$,
$$f(\theta|\gamma) = c(F^1(\theta^1|\gamma),..., F^n(\theta^n|\gamma)|\gamma) \prod_j f^j(\theta^j|\gamma).$$
Letting $z^j = F^j(\theta^j|\gamma)$ be the percentile of the buyer's valuation of the $j$th good and $v^j$ denote the mapping from percentiles and initial private information to eventual valuations, we have that $\theta^j = (F^j)^{-1}(z^j|\gamma) = v^j(z^j, \gamma)$. In this setting, the three sufficient assumptions can be stated as follows:
\begin{itemize}
    \item[A1'.] Commonly-Ordered': For each good $j$ and fixed $z^j \in [0, 1]$,
    $$v^j(z^j, \gamma)$$
    is increasing in $\gamma$. That is, for a fixed percentile, valuations are increasing in initial private information for all goods. 
    \item[A2'.] Sequential Regularity': Virtual value for good $j$ being increasing in $\gamma$ is equivalent to 
    $$v^j_{\gamma, \gamma}/v^j_{\gamma} \leq v^j_{\gamma, z^j}/v^j_{z^j}.$$
    That is, conditional on the same eventual valuation for good $j$, the buyer's initial private information $\gamma$ has a smaller marginal impact on the final valuation as $\gamma$ grows.
    
    Virtual value for good $j$ being increasing in $z^j$ is equivalent to 
    $$v^j_{\gamma, z^j} \leq 0.$$
    That is, buyers with higher initial private information $\gamma$ are (weakly) less impacted by the downstream shock $z^j$.
    \item[A3'.] Invariant Dependencies': The functions $C(z^1,...,z^n|\gamma), c(z^1,...,z^n|\gamma)$ can be written as $C(z^1,...,z^n), c(z^1,...,z^n)$. That is, the cumulative and probability distribution functions of the percentiles do not depend on $\gamma$.
\end{itemize}

The first two assumptions are standard regularity assumptions, adapted to the sequential screening setting, for each good. Invariant dependencies warrants further discussion. It is plausibly satisfied in many economic settings. If $\gamma$ captures the predictable portion of eventual demand, then residual uncertainty will generally be distributed similarly across individuals. Invariant dependencies also plays a role in simplifying the analysis. In particular, it rules out using correlation across goods as a screening device; $\gamma$ is only relevant insofar as it provides information about the marginal distribution of the buyer's valuations for each good.\footnote{However, this does not immediately imply that separate sales is optimal as the different goods are still linked via the monopolist's choice of how to optimally provide incentives for the buyer to truthfully report $\gamma$.} Compared to the general intractabilities associated with multidimensional screening, invariant dependencies enables analysis to go a long way. Even the literature on approximate optimality of certain mechanisms assumes independence between goods; invariant dependencies is a significant generalization beyond full independence.

\subsection{Examples Satisfying A1-A3}

\noindent \textbf{Example 1: Additive Independent Shock.} This would model retail consumers who are subject to stochastic cravings. Suppose 
$$\theta = u(\gamma) + \epsilon$$
where $\gamma$ follows any distribution with (weakly) decreasing inverse hazard rate, $u: \Gamma \to \mathbb{R}^n$ is deterministic and $\epsilon$ is a random variable with support $\mathbb{R}^n$. As long as $u^j(\gamma)$ is (weakly) increasing and (weakly) concave in $\gamma$ for each $j$ and $\epsilon$ is independent of $\gamma$, then assumptions A1-A3 are satisfied:
\begin{itemize}
    \item[A1'.] Since $\epsilon$ is independent of $\gamma$, the same quantile $z^j$ corresponds to the same realization of $\epsilon^j$ for all $\gamma.$ Then, $v^j(z^j, \gamma)$ being increasing in $\gamma$ for fixed $z^j$ is equivalent to $u^j(\gamma) + \epsilon^j$ being increasing in $\gamma$ for fixed $\epsilon^j$.
    \item[A2.] Sequential virtual value for good $j$ is 
    $$\theta^j - u_\gamma^j(\gamma) \frac{1-G(\gamma)}{g(\gamma)}$$
    which is clearly increasing in $\theta^j$. If $u$ is (weakly) concave and (weakly) increasing and $G$ has (weakly) decreasing inverse hazard rate, then sequential virtual value is increasing in $\gamma$ as well. 
    \item[A3'.] The copula is simply the copula for $\epsilon$ which is independent of $\gamma$.
\end{itemize}

Noise does not have to be perfectly additively separable: It could be that $\theta = u(\gamma) + \epsilon(\gamma)$. If the magnitude of $\epsilon$ does not vary too much as $\gamma$ varies, assumptions A1 and A2 will continue to hold. As long as the distribution of percentiles of $\epsilon$ does not vary at all as $\gamma$ varies, assumption A3 will continue to hold.

This parametrization of the model also demonstrates how multidimensional sequential screening contrasts with the usual multidimensional screening problem. Consider two extreme cases of when information arrives. In the first case, suppose all of the buyer's information arrives in $\gamma$ so $\epsilon = 0$ and the buyer's valuation is just given by $u(\gamma)$. If this is the case, the only restriction is that $u(\gamma)$ is increasing in each component. This immediately collapses the problem into a one-dimensional screening problem; in particular, the marginal gains from adding any good to a given bundle is independent of the original bundle. As such, the marginal revenue intuition of \cite{yang_2025_nested} applies and separately posting prices for each good \textit{induces} an allocation that features nested bundling, as buyers with higher values of $\gamma$ purchases larger (in the set order) bundles. In the second case, suppose all of the buyer's information arrives in $\epsilon$ so $u(\gamma)$ is constant in $\gamma$. In this case, the buyer has no (payoff-relevant) private information prior to contracting and the seller can achieve full surplus extraction by charging 
$$\E[\epsilon]{\sum_j \max\left\{u^j(\gamma) + \epsilon^j, 0\right\}}$$
upfront and posting a price for zero for each good.

\noindent \textbf{Example 2: Multiplicative Independent Shock.} Suppose
$$\theta = u(\gamma) \odot \epsilon$$
where $u: \Gamma \to \mathbb{R}_{\geq 0}^n$ is increasing and deterministic, $\epsilon$ is a random variable with support $\mathbb{R}_{\geq 0}^n$, and $a \odot b$ is the coordinate-wise product of two vectors. Suppose $G$ has decreasing inverse hazard rate and $u^j$ is log-concave and $\frac{u^j_\gamma(\underline{\gamma})}{u^j(\underline{\gamma})}\frac{1}{g(\underline{\gamma})} < 1$ for all $j$.\footnote{This states that the initial growth rate of $u$ cannot be too large nor can low values of $\gamma$ be too unlikely.} This would model airline travelers who stochastically may be unable to go on their flight (take the support of $\epsilon$ to be $\{0, 1\}^n$). Then, Assumptions A1-A3 are satisfied:
\begin{itemize}
    \item[A1'.] Since $\epsilon$ is independent of $\gamma$, the same quantile $z^j$ corresponds to the same realization of $\epsilon^j$ for all $\gamma.$ Then, $v^j(z^j, \gamma)$ being increasing in $\gamma$ for fixed $z^j$ is equivalent to $u^j(\gamma) \cdot \epsilon^j$ being increasing in $\gamma$ for fixed $\epsilon^j$. This is satisfied as $\epsilon^j \geq 0$ and $u^j(\gamma)$ is increasing.
    \item[A2'.] Sequential virtual value for good $j$ is 
    $$\theta^j \left[ 1 - \frac{u^j_\gamma(\gamma)}{u^j(\gamma)} \cdot \frac{1-G(\gamma)}{g(\gamma)} \right].$$
    Since $G$ has decreasing inverse hazard rate and $u^j$ is log-concave, $1 - \frac{u^j_\gamma(\gamma)}{u^j(\gamma)} \cdot \frac{1-G(\gamma)}{g(\gamma)}$ is increasing in $\gamma$. Since $u^j(\gamma), \epsilon^j$ are non-negative, sequential virtual values are (weakly) increasing in $\gamma$. Then, sequential virtual values are (weakly) increasing in $\theta$ if 
    $$1 - \frac{u^j_\gamma(\gamma)}{u^j(\gamma)} \cdot \frac{1-G(\gamma)}{g(\gamma)} \geq 0.$$
    As this term is increasing in $\gamma$, this holds if and only if $\frac{u^j_\gamma(\underline{\gamma})}{u^j(\underline{\gamma})}\frac{1}{g(\underline{\gamma})} < 1$.
    \item[A3'.] Since $u^j(\gamma)$ and $\epsilon$ are non-negative, the copula over $\theta$ is simply the copula for $\epsilon$ which is independent of $\gamma$.
\end{itemize}

\noindent \textbf{Example 3: Noisy Vertical Learning.} Suppose $\theta$ is normally distributed with mean $\mu$ and covariance matrix $\Sigma$. The buyer learns $\gamma = \alpha \cdot \theta + \epsilon$ where $\epsilon$ is normally distributed (a student t-distribution with any number of degrees of freedom also works similarly) with mean $0$ and variance $\sigma$. Suppose, similar to \cite{pernoud_2025_bundling},\footnote{\cite{pernoud_2025_bundling} do not consider noise in the buyer's signal but allow for elliptical distributions, which generalize normal distributions. Elliptical distributions would violate Invariant Dependencies as updating on $\gamma$ would then change the correlational structure. For example, suppose $\theta$ is drawn from either $N(0, I)$ or $N(0, kI)$ with equal (ex-ante) probability, where $I$ is the $n$-dimensional identity matrix and $k > 1$. Let $\gamma = \frac{1}{n} \sum_j \theta^j$ be the average value over all goods. As $\gamma$ grows large we can be confident that $\theta$ was drawn from $N(0, kI)$ instead of a mixture of the two distributions, changing the copula.} that learning is vertical; that is $\gamma$ is positively correlated with each dimension $\theta^j$.\footnote{This implicitly puts assumptions on $\Sigma$; one sufficient condition used in \cite{pernoud_2025_bundling} is to assume that all covariances are the same. Another possibility is to assume that all covariances are weakly positive; in general there cannot be too many covariances that are too negative.} Then, the distribution of $\theta$ conditional on $\gamma$ is normal with mean 
$$\mu^* = \mu + \frac{\text{Cov}(\theta, \gamma)}{\text{Var}(\gamma)} (\gamma - E[\gamma]) = \mu + \frac{\Sigma \alpha}{\alpha^T \Sigma \alpha + \sigma^2} (\gamma - \alpha^T \mu)$$
and covariance
$$\Sigma^* = \Sigma - \frac{\text{Cov}(\theta, \gamma) \text{Cov}(\theta, \gamma)^T}{\text{Var}(\gamma)} = \Sigma - (\Sigma \alpha) \left( \alpha^T \Sigma \alpha + \sigma^2 \right)^{-1} (\alpha^T \Sigma).$$
All three sufficiency assumptions are satisfied: A3 holds as $\Sigma^*$ is constant in $\gamma$. A1 holds since $\Sigma^*$ is constant in $\gamma$ and vertical learning implies that $\mu^*$ is increasing in $\gamma$. Finally, A2 holds since sequential virtual value for good $j$ is 
$$\theta^j - {\mu^*}_\gamma^j(\gamma) \frac{1-G(\gamma)}{g(\gamma)}$$
and $\mu^*$ is linearly increasing in $\gamma$ while the distribution $G$ has decreasing inverse hazard rate as $\gamma$ is normally distributed.

\section{Dynamic Decoupling}\label{decoupling}

Our main result is the following, establishing optimality of a separable mechanism: 
\begin{theorem}[Optimality of Separate Sales]\label{no_bundlding}
    Under Assumptions A1-A3, the optimal joint mechanism is to offer the optimal sequential screening mechanism for each good separately. It can be implemented via options pricing: A buyer reporting $\gamma$ secures a price for good $j$ of
    $$p^j(\gamma) = \min \left\{p: \phi^j(\gamma, p) \geq 0\right\}$$
    and pays an upfront cost of
    $$t_1(\gamma) = \sum_{j=1}^n \left[\int_{p^j(\gamma)}^{\bar{\theta}^j}1-F^j(\theta^j|\gamma) d\theta^j + \int_{\underline{\gamma}}^\gamma \int_{p^j(\gamma')}^{\bar{\theta}^j}F_\gamma^j(t|\gamma') dt d\gamma' \right].$$
\end{theorem}

Theorem \ref{no_bundlding} provides a novel rationalization of separate sales via screening intertemporally. Notably, this does not rely on simplicity, robustness, or other behavioral concerns. Practically, this means that instead of searching over the complex space of all menus over lotteries of bundles, monopolists that contract over time can instead search over the much smaller space of upfront fees for spot prices. This not only significantly simplifies the monopolist's problem but also leads to higher revenues. We next work through a sketch of the proof.

The analysis follows the Meyersonian approach of constructing an appropriate virtual valuation function before solving a relaxed maximization problem over allocation rules. There are two main steps: (1) formalizing the intuition that the monopolist can frontload surplus extraction to re-write the monopolist's problem (Lemma \ref{formulation}) and (2) showing that the resulting distortion to buyer information rents induces independent sales (Lemma \ref{add_sep}). Methodologically, the monopolist's objective must be formulated in utility space (unlike single-good sequential screening where the objective can be formulated directly in terms of the allocation rule). However, distortions to information rents induced by sequential screening allows us to return to optimizing over allocation rules (unlike static multidimensional screening, where it is intractable to switch from working in utility space to allocation rules). Verification of global incentive constraints (and other omitted details) are in the Appendix.

The first step: Frontloading surplus extraction. Let $u(\gamma, \theta) = \theta \cdot q(\gamma, \theta) - t_2(\gamma, \theta)$ be the buyer's utility net of $t_1$. For any $u(\gamma, \theta)$, applying the envelope theorem to period two gives that the allocation rule must be $q(\gamma, \theta) = \nabla_\theta u(\gamma, \theta)$. As such, total surplus $\theta \cdot \nabla_\theta u(\gamma, \theta)$ being equal to consumer surplus $u(\gamma, \theta)$ plus producer surplus $t_2(\gamma, \theta$ implies that period two transfers are the same as in standard multidimensional screening settings:
$$t_2(\gamma, \theta) = \theta \cdot \nabla_\theta u(\gamma, \theta) - u(\gamma, \theta).$$
Next, let 
$$U(\gamma'|\gamma) = \int_\Theta u(\gamma',\theta) dF(\theta|\gamma) - t_1(\gamma')$$
and be the interim utility a consumer with type $\gamma$ receives when reporting $\gamma'$ and
$$U(\gamma) = U(\gamma|\gamma) = \int_\Theta u(\gamma,\theta) dF(\theta|\gamma) - t_1(\gamma)$$
be the interim utility of a type $\gamma$ consumer when reporting truthfully. Since $\gamma$ is not directly payoff-relevant and only affects payoffs by influencing the distribution of $\theta$, incentive compatibility of truthfully revealing $\theta$ holding for all $\gamma$ implies that even after a deviation to $\gamma'$, it is still a best response to truthfully report $\theta$. Then, by incentive compatibility for revealing $\gamma$, we have $\gamma \in \argmax_{\gamma'} U(\gamma'|\gamma)$ so the Envelope Theorem applies and local incentive constraints implies
$$U_\gamma(\gamma) = U_\gamma(\gamma'=\gamma|\gamma) = \int_\Theta u(\gamma,\theta) f_\gamma(\theta|\gamma) d\theta.$$
By the Fundamental Theorem of Calculus, we have
$$U(\gamma) = U(\underline{\gamma}, \underline{\gamma}) + \int_{\underline{\gamma}}^\gamma U_\gamma(\gamma')d\gamma' = U(\underline{\gamma}, \underline{\gamma}) + \int_{\underline{\gamma}}^\gamma \int_\Theta u(\gamma',\theta) f_\gamma(\theta|\gamma') d\theta d\gamma'.$$
As usual, we can set $U(\underline{\gamma}, \underline{\gamma}) = 0$. Equating this with the direct computation of 
$$U(\gamma) = \int_\Theta u(\gamma,\theta) dF(\theta|\gamma) - t_1(\gamma)$$
as expected downstream utility minus transfers and solving for transfers gives
$$t_1(\gamma) = \int_\Theta u(\gamma,\theta) dF(\theta|\gamma) - \int_{\underline{\gamma}}^\gamma \int_\Theta u(\gamma',\theta) f_\gamma(\theta|\gamma') d\theta d\gamma'.$$
Thus, $t_1, t_2, q$ are all pinned down by $u$ so the monopolist only needs to maximize over $u$; the monopolist's objective function can be written as 
\begin{align*}
    \E[\gamma]{t_1(\gamma)} + \E[\gamma, \theta]{t_2(\gamma, \theta)} =& \int_\Gamma \left[\int_\Theta u(\gamma,\theta) dF(\theta|\gamma) - \int_{\underline{\gamma}}^\gamma \int_\Theta u(\gamma',\theta) f_\gamma(\theta|\gamma') d\theta d\gamma'\right] dG(\gamma) \\
    &+ \int_\Gamma \int_\Theta \left[\theta \cdot \nabla_\theta u(\gamma, \theta) - u(\gamma, \theta)\right] dF(\theta|\gamma) dG(\gamma).
\end{align*}
Crucially, the first term of $\E[\gamma]{t_1(\gamma)}$ and the second term of $\E[\gamma, \theta]{t_2(\gamma, \theta)}$ cancel out: Any information rents given to the consumer for their post-contractual information can be extracted, in expectation, before that information arrives. The remainder of the computation is standard, interchanging the two integrals with respect to $\gamma$. Expected transfers are
\begin{align*}
    & \int_\Gamma\left[ -\int_{\underline{\gamma}}^\gamma \int_\Theta u(\gamma',\theta) f_\gamma(\theta|\gamma') d\theta d\gamma' + \int_\Theta \theta \cdot \nabla_\theta u(\gamma, \theta) f(\theta|\gamma) d\theta \right] g(\gamma) d\gamma \\
    =& \int_\Gamma \int_\Theta \Big[-u(\gamma, \theta) f_\gamma(\theta|\gamma) (1-G(\gamma)) + \theta \cdot \nabla_\theta u(\gamma, \theta) f(\theta|\gamma) g(\gamma) \Big] d\theta d\gamma \\
    =& \E[\gamma, \theta]{\theta \cdot \nabla_\theta u(\gamma, \theta) -u(\gamma, \theta) \frac{f_\gamma(\theta|\gamma)}{f(\theta|\gamma)} \frac{1-G(\gamma)}{g(\gamma)}}.
\end{align*}
Putting the above computation together, monopolist's problem is as follows:

\begin{lemma}\label{formulation}
    The monopolist's problem can be written as
    \begin{align*}
        \max_{u(\gamma, \theta), t_1(\gamma)} \text{ } & \E[\gamma, \theta]{\theta \cdot \nabla_\theta u(\gamma, \theta) - u(\gamma, \theta) \frac{f_\gamma(\theta|\gamma)}{f(\theta|\gamma)}\frac{1-G(\gamma)}{g(\gamma)} } & \\
        \text{st.  } & \gamma \in \argmax_{\gamma'} \E[\theta|\gamma]{u(\gamma', \theta)} - t_1(\gamma') & IC1 \\
        & u(\gamma, \theta) \text{ is convex in } \theta; & IC2 \\
        & u(\gamma, \theta) \text{ is 1-Lipschitz in } \theta; & F \\
        & \E[\theta|\gamma]{u(\gamma, \theta)} - t_1(\gamma) \geq 0 \text{ for all } \gamma & IR
    \end{align*}
    where $u(\gamma, \theta) = \theta \cdot q(\gamma, \theta) - t_2(\gamma, \theta)$ is the buyer's utility net of $t_1$ and $\nabla_\theta u(\gamma, \theta)$ is the gradient of $u(\gamma, \theta)$ in $\theta$ at some fixed $\gamma$. Furthermore, a necessary condition for IC1 to hold is
    $$t_1(\gamma) = \int_\Theta u(\gamma,\theta) dF(\theta|\gamma) - \int_{\underline{\gamma}}^\gamma \int_\Theta u(\gamma',\theta) f_\gamma(\theta|\gamma') d\theta d\gamma'$$
    which is used to arrive at the objective function.
\end{lemma}

$IC1$ continues to pin down incentive compatibility for truthful reporting of $\gamma$ while $IC2$ captures incentive compatibility for truthful reporting of $\theta$. Next, $F$ is feasibility: The gradient of the buyer's post-contractual utility function $u$ is equal to the allocation rule which must lie between zero and one. Finally, $IR$ pins down interim individual rationality to accept the contract. Total surplus generated by the monopolist is $\E[\gamma, \theta]{\theta \cdot \nabla_\theta u(\gamma, \theta)}$ while a share $\E[\gamma, \theta]{u(\gamma, \theta) \frac{f_\gamma(\theta|\gamma)}{f(\theta|\gamma)}\frac{1-G(\gamma)}{g(\gamma)}}$ is given to the buyer as information rents. The Lemma establishes that the interim utility function $u$ pins down both period one and period two transfers and specifies how dynamics distorts monopolist profits.

Compared to the usual multi-good monopolist problem where the buyer receives their full information rents $u$, allowing the principal to frontload surplus extraction leads to period two information rents given to the buyer being extracted by period one transfers. As such, the only information rents the principal needs to pay the buyer are those to truthfully elicit their pre-contractual private information. These information rents are distorted by two factors. First, $\frac{f_\gamma(\theta|\gamma)}{f(\theta|\gamma)} = \frac{\partial}{\partial \gamma} \ln(f(\theta|\gamma)) = s(\gamma, \theta)$ is the score function measuring how much the log-likelihood of $\theta$ changes as $\gamma$ changes. This captures how likely $\theta$ came from one value of $\gamma$ versus another and thus governs how post-contractual utilities are passed through into information rents. The other $\frac{1-G(\gamma)}{g(\gamma)}$ term is the standard hazard rate; as usual, providing information rents to lower types leads to higher information rents to higher types. 

The second step: Exploiting this distortion to buyer information rents. The usual approach to multidimensional screening problems would be to apply the divergence theorem as in \cite{opttransport}, expressing $\nabla_\theta u(\gamma, \theta)$ in terms of $u(\gamma, \theta)$, and then applying duality.\footnote{There are two complications preventing us from using an optimal transport or duality (see \cite{kleiner2019}) approach to analyze the problem. First, as \cite{segal2014} notes, the consumer's expected interim utility function may be discontinuous in $\gamma$. Unlike continuity in $\theta$ being guaranteed by feasibility of the allocation rule, the principal may choose to give information rents to consumers with different values of $\gamma$ at different values of $\theta$. While is is possible to use an orthogonalization approach to transform the problem, making $\E[\theta|\gamma]{u(\gamma, \theta)}$ continuous in $\gamma$, it is unclear if $u(\gamma, \theta)$ itself can be made continuous in $\gamma$; or if doing so would be an useful exercise. Second, the usual single-crossing conditions that lead to monotonicity being a necessary and sufficient condition for implementability do not hold. We can view the problem as the principal designing a utility function $w$ where a buyer with private information $\gamma$ values $w$ at
$$\hat{U}(\gamma, w(\theta)) = \int_\Theta w(\theta) dF(\theta|\gamma).$$
Then, the usual characterizations of single crossing or monotonicity in $w$ cannot be applied as $w$ is a high-dimensional object. However, cyclic monotonicity as in \cite{rochet1987} is still a necessary and sufficient condition for implementability, but is difficult to operationalize.} However, due to the additional distortion of $f_\gamma(\theta|\gamma)$ stemming from only paying the buyer information rents scaling with the informativeness of their initial private information, it is possible to instead apply the divergence theorem to express $u(\gamma, \theta)$ in terms of $\nabla_\theta u(\gamma, \theta)$ in a simple way, especially so if invariant dependencies holds.

Formally, let $\mathbf{V}(\theta|\gamma)$ be a vector-valued function satisfying 
\begin{equation*}\label{V_cond}
    \nabla_\theta \cdot (\mathbf{V}(\theta|\gamma) f(\theta|\gamma)) = f_\gamma(\theta|\gamma).
\end{equation*}
Then, applying the Divergence Theorem (when boundary terms go to zero) gives
$$\int_\Theta u(\gamma, \theta) \left( \nabla_\theta \cdot (\mathbf{V}(\theta|\gamma) f(\theta|\gamma)) \right) d\theta = - \int_\Theta \nabla_\theta u(\gamma, \theta) \cdot (\mathbf{V}(\theta|\gamma) f(\theta|\gamma)) d\theta$$
so the objective can be written as 
\begin{equation}\label{eqn1}
    \E[\gamma, \theta]{\nabla_\theta u(\gamma, \theta) \cdot \left(\theta + \mathbf{V}(\theta|\gamma) \frac{1-G(\gamma)}{g(\gamma)}\right)}.
\end{equation}
If invariant dependencies holds, $\mathbf{V}(\theta|\gamma)$ has a simple form: The $j$th component of $\mathbf{V}$ is
\begin{equation*}\label{V_sol}
    \mathbf{V}^j(\theta|\gamma) = \frac{F^j_\gamma(\theta^j|\gamma)}{f^j(\theta^j|\gamma)}
\end{equation*}
and boundary terms do vanish.\footnote{At the boundary, the CDF is either zero or one regardless of $\gamma$ so $F_\gamma^j$ vanishes.} In the terminology of \cite{segal2014}, $V$ is a vector of impulse responses of $\theta$ to $\gamma$. Recalling that $\nabla_\theta u(\gamma, \theta) = q(\gamma, \theta)$ gives the following formulation:

\begin{lemma}\label{add_sep}
    Suppose A3: Invariant Dependencies holds. Then, the monopolist's problem can be written as:
    \begin{align*}
        \max_{q(\gamma, \theta)} \text{ } & \E[\gamma, \theta]{\sum_j q^j(\gamma, \theta) \phi^j(\gamma, \theta^j)} \\
        \text{st.  } & q(\gamma, \theta)  \text{ is cyclically monotone in } \theta; \\
        & q(\gamma, \theta) \in [0, 1]^n \text{ for all } \gamma, \theta; \\
        & IC1, IR.
    \end{align*}
    where
    $$\phi^j(\gamma, \theta^j) = \left(\theta^j + \frac{F_\gamma^j(\theta^j|\gamma)}{f^j(\theta^j|\gamma)}\frac{1-G(\gamma)}{g(\gamma)}\right)$$
    denotes virtual value for good $j$.
\end{lemma}

If $F$ has invariant dependencies then only the marginals matter for the objective function. Then, taking the standard approach and ignoring global incentive compatibility constraints, maximizing the objective pointwise (i.e. allocating the good whenever virtual value is positive) leads to an allocation rule $q(\gamma, \theta)$ for which the allocation rule for good $j$ does not depend on the valuation for any other good $k \neq j$. When, in addition to A3, assumptions A1 and A2 also hold, the solution to the relaxed problem also satisfies global incentive constraints: Assumption A2 gives that the optimal allocation rule is increasing in $\gamma$ (and $\theta$) while Assumption A1 implies that buyer interim utility has increasing differences in $\gamma, q$ so any increasing allocation rule satisfies global incentive constraints. Monotonicity in each dimension and independence across dimensions then implies cyclic monotonicity for the overall allocation rule. Since the relaxed problem in dimension $j$ exactly coincides with the relaxed problem if only good $j$ were present, the optimal mechanism is to run separate sequential screening mechanisms for each good.

Armed with Theorem \ref{no_bundlding}, we may revisit the illustrative example, where the proposed sequential screening mechanism indeed maximizes monopolist surplus. Recall the setting: There are two goods with valuations $\theta^1 = v-\epsilon, \theta^2 = v+\epsilon$ for $v$ uniform over $[0, 1/2]$ and $\epsilon$ uniform over $[-1/2, 1/2]$. The value $v$ is the buyer's pre-contractual private information and $\epsilon$ is the buyer's post-contractual private information. Thus, final valuations are both uniform over $[v-1/2, v+1/2]$ and perfectly anti-correlated. In the language of the model, this corresponds to $v = \gamma \sim U[0, 1/2]$ and $F^1(\theta^1|\gamma) = F^2(\theta^2|\gamma) = \max\{1-(\gamma+1/2-\theta), 0\}$. As such, virtual values in this setting are
$$\phi^i(\theta^i|\gamma) = \theta^i + \left(-\frac{1-2\gamma}{2}\right) = \theta^i - (1/2-v)$$
so selling whenever virtual value is non-negative corresponds to posting prices of $1/2-v$ for each good. Period one transfers can then be computed to be $2v^2$. The allocation rule can be plotted as follows:

\begin{figure}[h!]
\centering
    \begin{tikzpicture}[mysquare]
    \draw[thick] (0,-0.5) -- (0,0.5); 
    \draw[thick] (0,-0.5) -- (0.5,-0.5); 
    \draw (0,0.5) -- (0.5,0.5); 
    \draw (0.5,-0.5) -- (0.5,0.5); 

    \draw (0,0) node[left] {$\varepsilon \quad 0$} -- (0.01,0);
    \draw (0,0.5) node[left] {$1/2$} -- (0.01,0.5);
    \draw (0,-0.5) node[left] {$-1/2$} -- (0.01,-0.5);
    
    \node[below] at (0,-0.5) {$0$};
    \node[below] at (0.25,-0.5) {$1/4$};
    \node[below] at (0.5,-0.5) {$1/2 \ v$};

    \draw (0,-0.5) -- (0.5,0.5);
    \draw (0,0.5) -- (0.5,-0.5);

    \node at (0.08, 0) {$(0,0)$};
    \node at (0.25, 0.3) {$(0,1)$};
    \node at (0.25, -0.3) {$(1,0)$};
    \node at (0.42, 0) {$(1,1)$};
    \end{tikzpicture}
\end{figure}

\newpage

\section{Robustness to Violations of A1-A3} \label{relax}

\subsection{Unknown Copulas} \label{relax_A3}

There are two immediate ways to relax invariant dependencies while keeping the spirit of the assumption. To isolate screening marginals via $\gamma$ and screening via the joint distribution, suppose $\gamma$ and the copula $c$ are both the buyers' private information. Thus if the monopolist elicits $\gamma$ they only know the buyer's marginal distribution of values for each good.

\begin{proposition}\label{weaker_A3}
    Suppose that either:
    \begin{enumerate}
        \item The distribution of $\gamma$ is statistically independent from the distribution of $c$;
        \item The monopolist's objective is to maximize profits over the worst-case distribution of $c$.
    \end{enumerate}
    Then, the mechanism identified in Theorem \ref{no_bundlding} remains optimal. 
\end{proposition}

While optimality of the mechanism identified in Theorem \ref{no_bundlding} relies on invariant dependencies, the allocation rule, transfer rule, and resulting monopolist profits do not depend on the copula $c$. As such, the same mechanism is well-defined even if $c$ is not known to the monopolist.\footnote{Alternatively, the designer can simply offer a mechanism that is invariant to the buyer's report of $c$ so incentive compatibility is trivially satisfied.} Monopolist profits are constant in the copula since goods are sold separately in the optimal mechanism anyway. The proof of Proposition \ref{weaker_A3} is straightforward.

\begin{proof}
    Suppose condition one holds. Consider the relaxed problem where the monopolist knows $c$. Then, invariant dependencies holds conditional on knowing the copula. Furthermore, since $c$ and $\gamma$ are statistically independent, knowledge of $c$ does not give the monopolist any knowledge of $\gamma$. As such, the distribution of $\gamma$ remains unchanged so the optimal mechanism copula-by-copula is the mechanism identified in Theorem \ref{no_bundlding}. This upper bound to the monopolist's profit is achieved by the mechanism identified in Theorem \ref{no_bundlding}.

    Next, suppose condition two holds. One potential distribution over $c$ is a point mass at some copula. This then satisfies invariant dependencies so the mechanism identified in Theorem \ref{no_bundlding} is optimal; the profit of this mechanism is an upper bound for the value of the monopolist's profit. The mechanism in Theorem \ref{no_bundlding} achieves the upper bound for all distributions over copulas nature may choose.
\end{proof}

What happens if we go further and break the spirit of A3? Without invariant dependencies, the monopolist can use heterogeneity in correlational structures as a tool to screen for valuations itself. In this case, (nonlinear) bundling may once again be part of the monopolist's optimal mechanism. Such screening is seen in practice: Consider vacations on cruise ships. Cruise lines offer different goods at different qualities such as better cabins, on-board internet, alcoholic beverages, specialty dining, and onshore excursions. When buying a low or medium-tier cabin, there are separate upgrade prices to purchase the aforementioned add-ons. Furthermore, these upgrade prices are independent of cabin quality; whether or not the cabin has a balcony is priced separately from whether or not it comes with internet. However, for top-tier cabins (for example, the ``family villa'' on Norwegian Cruise Lines), each of the add-ons are bundled with the cabin purchase itself. This is rationalized by a situation where consumers with lower values of pre-contractual information have valuations for add-ons that are independent of one another (valuations for add-ons are essentially noise), but consumers with higher values of initial information have correlated valuations for add-ons (some premia for a vacation being ``all-inclusive'').

To illustrate screening valuations via correlations and the analytical complications that arise without invariant dependencies, consider the illustrative example with the following change: Let valuations remain perfectly anti-correlated for $\gamma < 1/4$ but have valuations be perfectly correlated for $\gamma \geq 1/4$:
$$\theta^1 = v+\epsilon, \theta^2 = v-\epsilon \text{ if } v < 1/4$$
$$\theta^1 = v+\epsilon, \theta^2 = v+\epsilon \text{ if } v \geq 1/4$$
where 
$$v \sim U[0, 1/2]; \epsilon \sim U[-1/2, 1/2].$$
This leads incentive constraints to discontinuously change between the perfect correlation and perfect anti-correlation regimes. Even while keeping the same induced allocation rule as the original setting, the monopolist can increase payments to take advantage of the change by imposing an additional surcharge for purchasing the bundle. The monopolist can do even better by more aggressively allocating the buyer's more-preferred good (but not the bundle) to buyers in the anti-correlation regime.

Recall the previous mechanism of charging $2v^2$ upfront to secure strike prices of $1/2-v$. Note that if $v < 1/4$ no consumer buys the grand bundle while if $v \geq 1/4$ consumers only buy the grand bundle. Thus, while strike prices of $1/2-v$ implements this allocation, the monopolist can do better via the following mechanism, charging a premia for the bundle:
\begin{enumerate}
    \item If $v < 1/4$: Pay $2v^2$ upfront for the right to purchase one (but not both) goods at a price of $1/2-v$.
    \item If $v \geq 1/4$: Pay $2v^2+1/8$ upfront for the right to purchase the bundle of goods at a price of $1-2v$.
\end{enumerate}
This induces the following allocation rule:

\begin{figure}[h]
    \centering
    \begin{tikzpicture}[mysquare]
    \definecolor{myblue}{RGB}{45, 62, 144}
    \definecolor{myred}{RGB}{145, 30, 30}

    \draw[thick] (0,-0.5) -- (0,0.5); 
    \draw[thick] (0,-0.5) -- (0.5,-0.5); 
    \draw (0,0.5) -- (0.5,0.5); 
    \draw (0.5,-0.5) -- (0.5,0.5); 

    \draw (0,0) node[left] {$\varepsilon \quad 0$} -- (0.01,0);
    \draw (0,0.5) node[left] {$1/2$} -- (0.01,0.5);
    \draw (0,-0.5) node[left] {$-1/2$} -- (0.01,-0.5);
    
    \node[below] at (0,-0.5) {$0$};
    \node[below] at (0.25,-0.5) {$1/4$};
    \node[below] at (0.5,-0.5) {$1/2 \ v$};

    
    \draw[myblue, thick] (0,0.5) -- (0.25,0);
    \draw[myblue, thick] (0,-0.5) -- (0.25,0);
    \draw[myblue, thick] (0.25,-0.5) -- (0.25,0);
    
    \draw[myred, thick] (0.25,0.5) -- (0.25,0);
    \draw[myred, thick] (0.25,0) -- (0.5,-0.5);

    
    \node[myblue] at (0.175, 0.325) {$(0,1)$};
    \node[myblue] at (0.1, 0) {$(0,0)$};
    \node[myblue] at (0.175, -0.325) {$(1,0)$};
    
    \node[myred] at (0.4, 0) {$(1,1)$};
    \node[myred] at (0.325, -0.325) {$(0,0)$};
    \end{tikzpicture}
\end{figure}

The only new incentive constraint introduced is along the border of $v = 1/4$. Clearly, there will not be any upward deviations as a buyer with $v = 1/4-\epsilon$ cannot make any surplus from buying their less-preferred good at a price of $1/2-1/4 = 1/4$. A buyer with $v = 1/4$ receives surplus of 
$$\underbrace{2 \cdot \frac{1}{2} \cdot (1/4+1/4)^2}_{\text{expected downstream gains}} - \underbrace{2(1/4)^2}_{\text{upfront payment}} - \underbrace{1/8}_{\text{premia}} = 1/8 - 1/8 = 0$$
from being truthful. If they report $1/4-\epsilon$, their profit, up to $\epsilon$ terms, is:
$$\underbrace{\frac{1}{2}(1/4+1/4)^2}_{\text{single good}} - \underbrace{2(1/4)^2}_{\text{upfront payment}} = 0.$$
As such, this alternative mechanism satisfies incentive and individual rationality constraints. The monopolist is able to charge an extra $1/8$ with probability $1/2$ (i.e. whenever $v > 1/4$) and their profit is thus (recall that profit from the original mechanism was $1/3$):
$$\frac{1}{3} + \frac{1}{2} \cdot \frac{1}{8} = \frac{19}{48}.$$
Total surplus is unchanged at $1/2$ so consumer surplus is decreased (by exactly the $1/16$ that the monopolist can now additionally extract):
$$\frac{1}{2}-\frac{19}{48} = \frac{5}{48} < \frac{8}{48} = \frac{1}{6}.$$
The intuition is as follows: Since the correlational structure changes, the set of options in the monopolist's menu also changes. This allows the monopolist to ``reset'' information rents at $v = 1/4$. Buyers with $v = 0$ and $v = 1/4$ both receive zero surplus in expectation. 

What is the optimal mechanism? To illustrate the forces involved, let us return to the general setting for now. Suppose $\Gamma$ can be partitioned into
$$\Gamma = [\underline{\gamma}, \overline{\gamma}] = [\underline{\gamma}, \gamma_1] \cup [\gamma_1, \overline{\gamma}]$$
such that invariant dependencies holds over $[\underline{\gamma}, \gamma_1]$ and $\cup [\gamma_1, \overline{\gamma}]$ separately. In an abuse of notation, the mechanism will be doubly-defined for $\gamma = \gamma_1$ since it is in both partitions; this is measure zero and is done to avoid infinitesimal terms. In this setting, the problem can be written as follows: The monopolist seeks to maximize 
\begin{align*}
    & \int_{[\underline{\gamma}, \gamma_1]} \left[\int_\Theta \theta \cdot \nabla_\theta u(\gamma, \theta) d\theta - \int_{\underline{\gamma}}^\gamma \int_\Theta u(\gamma', \theta) f_\gamma(\theta|\gamma') d\theta d\gamma' \right] g(\gamma) d(\gamma) \\
    & + \int_{[\gamma_1, \overline{\gamma}]} \left[\int_\Theta \theta \cdot \nabla_\theta u(\gamma, \theta) d\theta - U_1 - \int_{\gamma_1}^\gamma \int_\Theta u(\gamma', \theta) f_\gamma(\theta|\gamma') d\theta d\gamma' \right] g(\gamma) d(\gamma)
\end{align*}
where $u(\gamma, \theta)$ is the utility function the monopolist designs and $U_1$ is the utility of an agent with $\gamma = \gamma_1$ at the threshold. For buyers with $\gamma \in [\underline{\gamma}, \gamma_1]$, this term is the same as before. However, for buyers with $\gamma \in [\gamma_1, \overline{\gamma}]$, the Fundamental Theorem of Calculus is applied with $\gamma_1$ at the lowest point instead of $\underline{\gamma}$; this leads to baseline utility being $U_1$ instead of $0$ but the integral starting at $\gamma_1$ instead of $\underline{\gamma}$. The new constraints are incentive compatibility and individual rationality at the boundary ($U_1 \geq 0$). Applying the tools developed in this paper to each interval separately gives the following objective:
\begin{align*}
    & \E[{\gamma \in [\underline{\gamma}, \gamma_1], \theta}]{\sum_j q^j(\gamma, \theta) \left(\theta^j + \frac{F_\gamma(\theta^j|\gamma)}{f^j(\theta^j|\gamma)}\frac{G(\gamma_1)-G(\gamma)}{g(\gamma)}\right)} \\
    & + \E[{\gamma \in [\gamma_1, \overline{\gamma}], \theta}]{\sum_j q^j(\gamma, \theta) \left(\theta^j + \frac{F_\gamma(\theta^j|\gamma)}{f^j(\theta^j|\gamma)}\frac{1-G(\gamma)}{g(\gamma)}\right)}-U_1[1-G(\gamma_1)].
\end{align*}
There are two forces that go in opposite directions. The standard ``no distortions at the top'' intuition manifests itself in allocating efficiently to the highest type in $[\underline{\gamma}, \gamma_1]$ as their virtual value for good $j$ is 
$$\theta^j + \frac{F_\gamma(\theta^j|\gamma)}{f^j(\theta^j|\gamma)}\frac{G(\gamma_1)-G(\gamma = \gamma_1)}{g(\gamma)} = \theta^j.$$
This suggests that the monopolist should allocate more of the good than before to buyers in $[\underline{\gamma}, \gamma_1]$. On the other hand, if too much of the good is allocated to agents at the high end of $[\underline{\gamma}, \gamma_1]$, the value of $U_1$ must increase to deter downward deviations. This alternative force suggests that the monopolist should allocate less of the good than before to buyers in $[\underline{\gamma}, \gamma_1]$. How these forces are balanced also depends on how the distribution of $\theta$ discontinuously changes at $\gamma_1$. Furthermore, there are multiple mechanisms where both individual rationality and incentive compatibility at the border are binding; the difficulties associated with multidimensional screening come back in this case. Another takeaway is that the optimal mechanism for the upper partition remains unchanged. 

Return to the illustrative example with these intuitions: The monopolist would like to allocate more of the good to buyers in $[0, 1/4]$ while preserving giving the lowest type in $[1/4, 0]$ zero information rents. Using $G(1/4)$ instead of $G(1/2) = 1$ in the expression for virtual values for buyers in $[0, 1/4]$ gives their virtual value to be 
$$\theta^i + \left(\frac{-1/2 - 2\gamma}{2}\right) = \theta^i - (1/4-v)$$
which suggests strike prices of $1/4-v$. However, buyers in $[1/4, 1/2]$ value the option of purchasing a bundle much more than buyers in $[0, 1/4]$ since their values are perfectly correlated (instead of being perfectly anti-correlated). This suggests the following mechanism (period one transfers are calculated using the standard formula for transfers):
\begin{enumerate}
    \item If $0 \leq v \leq 1/8$: Pay $2v^2 + \frac{1}{2}v + \frac{1}{16}$ to secure a strike price of $1/4-v$ for one good.
    \item If $1/8 \leq v \leq 1/4$: Pay $\frac{9}{32}$ upfront to claim exactly one of the goods for free.
    \item If $1/4 \leq v$: Pay $2v^2+1/8$ upfront for the right to purchase the bundle of goods at a price of $1-2v$.
\end{enumerate}

Case one reflects the pointwise optimal mechanism. Case two is the optimal mechanism subject to not selling a bundle. Note that at $v = 1/8$, strike prices are $1/4-1/8 = 1/8$ so the consumer will always purchase one of the goods. As such, it is without loss to charge them the $1/8$ upfront; this is added to the standard upfront fee to arrive at $9/32$. Charging them upfront instead of downstream helps with preventing downward deviations from buyers in $[1/4, 1/2]$: These buyers will not purchase either good at a price of $1/8$ with some probability so frontloading the transfer hurts their deviation payoff. Finally, case three is the optimal mechanism from before with a bundle premia to set the utility of the lowest type in $[1/4, 1/2]$ to be equal zero. This induces the following allocation rule:
\begin{figure}[h]
    \centering
    \begin{tikzpicture}[mysquare]
    \definecolor{myblue}{RGB}{45, 62, 144}
    \definecolor{myred}{RGB}{145, 30, 30}
    \definecolor{mypurple}{RGB}{102, 0, 153}

    \draw[thick] (0,-0.5) -- (0,0.5); 
    \draw[thick] (0,-0.5) -- (0.5,-0.5); 
    \draw (0,0.5) -- (0.5,0.5); 
    \draw (0.5,-0.5) -- (0.5,0.5); 

    \draw (0,0) node[left] {$\varepsilon \quad 0$} -- (0.01,0);
    \draw (0,0.5) node[left] {$1/2$} -- (0.01,0.5);
    \draw (0,-0.5) node[left] {$-1/2$} -- (0.01,-0.5);
    
    \node[below] at (0,-0.5) {$0$};
    \node[below] at (0.25,-0.5) {$1/4$};
    \node[below] at (0.5,-0.5) {$1/2 \ v$};
    
    
    \draw[mypurple, thick] (0, 0.5) -- (0.125, 0);   
    \draw[mypurple, thick] (0, -0.5) -- (0.125, 0);  
    \draw[mypurple, thick] (0.125, 0) -- (0.25, 0);  
    \draw[myblue, thick] (0.25,-0.5) -- (0.25,0);
    
    \draw[myred, thick] (0.25,0.5) -- (0.25,0);
    \draw[myred, thick] (0.25,0) -- (0.5,-0.5);

    
    \node[myblue] at (0.05, 0) {$(0,0)$};
    \node[myblue] at (0.12, 0.3) {$(0,1)$};
    \node[myblue] at (0.12, -0.3) {$(1,0)$};
    
    \node[myred] at (0.375, 0.2) {$(1,1)$};
    \node[myred] at (0.32, -0.3) {$(0,0)$};

    \end{tikzpicture}
\end{figure}

Finally, we verify incentive constraints around $1/4$. A buyer in $[0, 1/4]$ deviating upwards will pay $2(1/4)^2+1/8 = 1/4$ to have the option to pay $1-2(1/4) = 1/2$ to receive a bundle they value at $(1/4-\epsilon) + (1/4+\epsilon) = 1/4$. Thus their expected payoff is $-1/4$ but their payoff from being truthful is strictly positive. A buyer in $[1/4, 1/2]$ deviating downward will pay $9/32$ to get an item. The expected value of an item is 
$$\E[{\epsilon\sim U[-1/2, 1/2]}]{\max{1/4+\epsilon, 0}} = 9/32$$
so their deviation payoff is $9/32-9/32 = 0$ and thus not profitable. Since $q$ is increasing in $\gamma$ separately over $[0, 1/4]$ and $[1/4, 1/2]$, larger deviations between $[0, 1/4]$ and $[1/4, 1/2]$ can be ruled out as well. As such, this mechanism is incentive compatible. Monopolist profits are 
$$\frac{157}{384} > \frac{152}{384} = \frac{19}{48}$$
while total surplus is 
$$\frac{71}{128} > \frac{64}{128} = \frac{1}{2}$$
and consumer surplus is 
$$\frac{7}{48} > \frac{5}{48}.$$
As such, this mechanism, which further exploits the change in correlational structure by forcing low-value buyers to only acquire a single good, increases both firm and consumer surplus.

It remains an open question if this is the globally optimal mechanism. Both individual rationality and incentive compatibility constraints bind at the boundary of $v = 1/4$. It generally is impossible to give individuals in $[0, 1/4]$ more of the good without keeping information rents of the lowest type in $[1/4, 1/2]$ at zero. 

\subsection{Relaxing A2 or A1}

Assumption A2 (Sequential Regularity) establishes that the pointwise maximizer of the relaxed problem is an allocation rule that is increasing in both $\gamma$ and $\theta^j$. Allocation rules being increasing in $\theta^j$ is necessary and sufficient (given additive separable sales) for period two incentive constraints to hold. Assumption A1 (Common FOSD Order) implies that any allocation rule being weakly increasing in $\gamma$ satisfies global incentive constraints. 

There are two dimensions to relax this pair of constraints: less regularity in $\gamma$ or less regularity in $\theta$. Without regularity in $\gamma$, different goods become endogenously linked via which good(s) are utilized to provide period one incentives. For example, suppose that for some subset of goods, either (1) higher values of $\gamma$ led to FOSD-lower distributions of values or (2) the pointwise solution to the relaxed problem was decreasing in $\gamma$. Either of these effects would provide an incentive for the buyer to shade their report of $\gamma$ to collect higher surplus from transacting on that subset of goods. However, incentive compatibility for the goods which satisfy A1 and A2 implies that this deviation leads the buyer to collect lower surplus from the remainder of the goods. If violations of A1 or A2 were sufficiently small on some subset of goods, the pointwise solution remains implementable. Otherwise, the monopolist may need to optimize over which dimensions are used to provide incentives.

What about regularity in $\theta$? A natural question would be when the optimal mechanism still features no non-linear bundling and ironing separately in each dimension is sufficient.

\begin{proposition}[Separate Ironing]\label{ironing}
    Suppose valuations are independent across goods and 
    $$\theta^j f^j(\theta^j|\gamma) g(\gamma) + F^j_\gamma(\theta^j|\gamma)(1-G(\gamma))$$
    is increasing in $\gamma$ for each $j$ (but not necessarily in $\theta^j$). Then, separately ironing each $\phi^j(\gamma, \theta^j)$ to form an ironed (in $\theta^j$) virtual value produces the optimal mechanism.
\end{proposition}

Independence is necessary for the problem to be strongly decomposable in each dimension. Without independence, a cyclically monotone $q(\gamma, \theta)$ may induce interim allocations 
$$\hat{q}^j(\gamma, \theta^j) = \E[\theta^{-j}]{q(\gamma, \theta)}$$
that are not monotone in $\theta^j$ based on how valuations are coupled; this holds even if assumption A3 is in effect. With independence, every cyclically monotone $q$ induces monotone allocation rules in each dimension so it is without loss to set $q^j$ solely as a function of $\gamma, \theta^j$. The other assumption supposes regularity on virtual values weighted by the probability of realizing $(\gamma, \theta)$ pairs. Under this condition, the optimization problem in each dimension has increasing differences in $q, \gamma$ so higher values of $\gamma$ yield pointwise-higher allocation rules, regardless of how ironing is done. 

\section{Extensions}\label{extensions}

To test the limits of how powerful frontloading surplus extraction is, we compare the model studied in this paper to two other benchmarks. First, one additional period is all the monopolist needs: If values are independent and private information for each good comes over time (as in \cite{segal2014}), the monopolist cannot utilize future allocations to extract additional surplus. Here, we need independence and not just invariant dependencies since if information comes over time, any dependencies can give the monopolist more power, thus stacking the deck too much in the monopolist's favor. Second, we suppose that the monopolist is able to directly observe an orthogonalized signal of the buyer's post contractual information (as in \cite{EsoSzentes2017}). Once again, the monopolist cannot benefit, providing another interpretation of frontloading surplus extraction allowing the monopolist to avoid paying \textit{any} post-contractual information rents. Third, we consider the case where there are $n$ items, each in unit supply, to be auctioned as in \cite{EsoSzentes2007}. In this case, the optimal auction for the $n$ items is to host separate sequential screening auctions for each good. 

\subsection{Additional Periods}

If the optimal mechanism is to sell goods independently, do dynamics have bite? The answer is no. Consider an alternative environment where instead of buying $n$ products simultaneously, there were $n+1$ periods. In the first period, the buyer draws $\gamma$ and the monopolist writes a contract. In each subsequent period $i$ the buyer's valuation $\theta^i$ is drawn according to $F^i(\theta^i|\gamma)$ and good $i$ is sold via some mechanism. Call this the sequential case. In general, profits from the original problem of selling simultaneously are upper bounded by profits from the alternative case of selling sequentially: When selling simultaneously, the buyer draws all of their valuations at once and is able to better coordinate joint deviations across goods. As such, any incentive compatible mechanism in the original problem is also incentive compatible in the sequential case, implying that the value of the sequential case is lower-bounded by the value of the original problem. However, this bound is tight whenever values are independent: Endowing the monopolist with additional power to screen intertemporally does not increase profits. Since all information rents are pulled forward into period one anyways, additional periods beyond period two are inconsequential. Note that this result is not a direct consequence of \cite{EsoSzentes2017}: Dynamic irrelevance in their setting only pertains to not needing to pay information rents to elicit future information so the same allocation rule leads to the same profit. I contribute by showing that even when expanding the space of implementable allocation rules, the principal does not do better when screening additionally through time. As such, the role of $\gamma$ (learning about learning) is qualitatively different than $\theta$ (learning about valuations).

\begin{proposition}[Irrelevance of Additional Periods]\label{seq_irrelevance}
    If values are independent, then the optimal allocation rule and profit from selling simultaneously coincides with the optimal allocation rule and profit from selling sequentially.
\end{proposition}

If there is dependence and different dimensions are drawn sequentially, however, different components of $\theta$ leads to both learning about learning and learning about valuations. The proof of Proposition \ref{seq_irrelevance} compares the solution to a multidimensional sequential screening problem to the optimal mechanism identified in \cite{segal2014}; absent independence, the optimal mechanism in the sequential case is difficult to derive.

\subsection{Observed Orthogonalization}\label{observed}

The standard approach to solving dynamic contracting problems is via orthogonalization, transforming the buyer's future information so that it is independent of their current information. In this case, \cite{EsoSzentes2017} shows dynamic irrelevance in a different sense: For any implementable allocation rule, making the buyer's future orthogonalization information publicly observable does not change the monopolist's profits. Does that result hold in the multidimensional setting and can it be operationalized?

To model the orthogonalization approach and benchmark in the multidimensional setting, let $v(\gamma, z)$ be some function such that for $z \sim U([0,1]^n)$ we have $v(\gamma, z) \sim F(\theta|\gamma)$ so $\theta = v(\gamma, z)$.\footnote{This is similar a copula but dependencies between dimensions are captured in $v$ itself while $z$ is fully uniform instead of only having uniform marginals. A copula, on the other hand, captures dependencies via the coupling function $C$.} One such mapping is
$$v(\gamma, z) = \left((F^1)^{-1}(z^1|\gamma), (F^2)^{-1}(z^2|\theta^1, \gamma),...,(F^n)^{-1}(z^n|\theta^1,...,\theta^{n-1},\gamma)\right).$$
In this setting (call it the relaxed problem), the monopolist's problem is to choose $\hat{q}: \Gamma \times [0, 1]^n \to [0, 1]^n, \hat{t}: \Gamma \times [0, 1]^n \to \mathbb{R}$. Since screening essentially only happens once, only one transfer rule $\hat{t}$ is needed, instead of $t_1, t_2$ as before. As such, the monopolist's problem is:
\begin{align*}
    \max_{\hat{t}, \hat{q}} & \: \E[\gamma]{\hat{t}(\gamma)} \\
    \text{s.t.} & \: \gamma \in \argmax_{\gamma'} \E[z]{\hat{q}(\gamma', z) \cdot v(\gamma, z) - \hat{t}(\gamma')}; & IC' \\
    & \: \E[z]{\hat{q}(\gamma, z) \cdot v(\gamma, z)} - \hat{t}(\gamma) \geq 0 \text{ for all } \gamma \in \Gamma & IR'
\end{align*}

Two notes about this formulation. First, it would also be possible to write down a model where $\hat{t}$ also depends on the realized $z$, but by the same pushforward intuition as earlier, we can take $\hat{t}(\gamma) = \E[z]{\hat{t}(\gamma, z)}.$ Second, it is inconsequential if $z$ were observed before or after contracting; $IC'$ could be replaced with a pointwise ($z$-by-$z$) incentive constraint instead. This case could be interpreted as a static multidimensional screening problem, but the buyer's private information can be projected down to a one-dimensional parameter. 

By similar reasoning as Lemma \ref{formulation}, we are able to recover the irrelevance result of \cite{EsoSzentes2017}, even when there is dependence between dimensions (recall that Lemma \ref{formulation} did not require independence either). However, Lemma \ref{formulation} clarifies the mechanism by which post-contractual information rents need not be paid: they have already been extracted in period one. On the other hand, the reasoning of \cite{EsoSzentes2017} directly jumps to the case where no information rents need to be paid.

\begin{proposition}[No Post-Contractual Information Rents]\label{same_rents}
    Suppose $q$ is implementable in the original problem. Then, the monopolist's profits from $q$ in the original problem are equal to their profits in the relaxed problem.
\end{proposition}

However, like the original versus sequential case, Proposition \ref{same_rents} does not imply that the original and relaxed problems are the same since implementability is an issue; more mechanisms can be implemented in the relaxed problem. When the distribution over values satisfies invariant dependencies, this gap disappears and the solution to the relaxed problem coincides to the solution of the original problem. 

\begin{proposition}[Relaxed Solution is Implementable]\label{yes_orthog}
    If assumptions A1-A3 are satisfied, then the solution of the original problem coincides with the solution of the relaxed problem.
\end{proposition}

However, if invariant dependencies is violated, solutions to the relaxed problem are not implementable in the original problem and \textit{cyclic} monotonicity becomes the key binding constraint. In the sequential version of the problem, cyclic monotonicity does not have any bite as contracting is done one dimension at a time.

\subsection{Downstream Auction}

Suppose there were $B$ bidders indexed by subscripts with generic element $b$. Each bidder observes $\gamma_b \sim G_b(\gamma)$ before contracting with the auctioneer and later observes $\theta_b \sim F_b(\theta_b|\gamma_b)$.\footnote{As in \cite{EsoSzentes2007}, we could also allow the auctioneer to observe post-contractual information via the multidimensional orthogonalization developed in Section \ref{observed}.} The auctioneer chooses an allocation rule 
$$q: \prod_b \Gamma_b \times \prod_b \Theta_b \to [0, 1]^{N \times B}$$
mapping all private information to quantities of each good for each bidder. Following the same processes as this paper and \cite{EsoSzentes2007}, the auctioneer's problem can be written as:
\begin{align*}
    \max_{q(\{\gamma_b, \theta_b\}_b)} \text{ } & \E[\{\gamma_b, \theta_b\}_b]{\sum_j q^j_b(\gamma_b, \theta_b) \phi^j_b(\gamma_b, \theta^j_b)} \\
    \text{st.  } & \E[\{\gamma_\beta, \theta_\beta\}_{\beta \neq b}]{q_b(\gamma_b, \theta_b)}  \text{ is cyclically monotone in } \theta_b \text{ for each } b; \\
    & q_b(\gamma_b, \theta_b) \in [0, 1]^n \text{ for all } \gamma, \theta, b; \\
    & \sum_b q^j_b(\{\gamma_b, \theta_b\}_b) \leq 1 \text{ for each } j.\\
    & IC1, IR.
\end{align*}
where
$$\phi^j_b(\gamma_b, \theta^j_b) = \left(\theta^j_b + \frac{F_{\gamma, b}^j(\theta^j_b|\gamma_b)}{f^j_b(\theta^j_b|\gamma_b)}\frac{1-G_b(\gamma_b)}{g_b(\gamma_b)}\right)$$
denotes bidder $b$'s virtual value for good $j$.

The pointwise solution to the auctioneer's problem is to set 
$$q^j_\beta(\{\gamma_b, \theta_b\}_b) =  \begin{cases}
1 & \text{if}\; \phi^j_\beta(\gamma_\beta, \theta^j_\beta) \geq 0, \phi^j_\beta(\gamma_\beta, \theta^j_\beta) \geq \phi^j_b(\gamma_b, \theta^j_b) \text{ for all } b \\
0 & \text{else}.
\end{cases}$$
That is, sell each good to the bidder with the highest non-negative virtual value and do not sell otherwise. Since virtual values are monotone in $\gamma_b$ for each bidder $b$ and good $j$ (by A2), this allocation rule is monotone in $\gamma_b, \theta^j_b$ and satisfies global period one incentive constraints (by A1). Since virtual values are monotone in $\theta^j_b$ (by A2) and the allocation rule is additive across goods, cyclic monotonicity is satisfied. Since these properties hold pointwise for any realization of other bidders' private information, taking expectations over $\theta_{-b}$ gives that all constraints are satisfied in the aggregate. This allocation rule can be implemented by running a handicap auction for each good independently of other goods. 

\section{Discussion}\label{discussion}

Endowing a principal the ability to contract intertemporally enables frontloading surplus extraction. This distorts the information rents which need to be paid to the buyer in favor of the monopolist, increasing profits and simplifying the multidimensional screening problem when distributions satisfy invariant dependencies. In this case, the distortion of information rents enables the monopolist's problem to be converted into one that is separable across dimensions; this transformation solves a probability continuity equation and corresponds to a statistical test of the buyer's reported pre-contractual information being separable across dimensions. When this is the case, the optimal mechanism sells goods separately, in contrast to the complicated non-linear bundling that usually arise in solutions to multidimensional screening problems. However, these solutions are not often used. On the other hand, the optimal mechanism developed in this paper parallels schemes often employed in practice. 

There are several avenues for further investigation. First, offering the optimal sequential screening mechanism independently for each good is still a \textit{feasible} mechanism even when invariant dependencies does not hold: Are there more general conditions for which it is optimal? Away from the invariant dependencies case, the function $\mathbf{V}$ may no longer be the same. However, a more complicated $\mathbf{V}$ may still be independent across dimensions, leading to a different pointwise maximization solution; even if the pointwise maximum is not implementable due to violations of cyclic monotonicity, it may still be the case that contracting independently across goods remains optimal. 

Second is analyzing the welfare effects of endowing the principal the ability to screen through time. Compared to the case where the monopolist can only contract after valuations themselves are known to the consumer, the monopolist always makes more profits in the sequential case (ignoring the buyer's pre-contractual private information and offering the optimal static multidimensional screening mechanism is always feasible). On the other hand, consumer information rents are distorted downwards but since it costs the monopolist less to provide information rents, realized information rents (net of distortions) may be larger. Investigating both how total surplus changes and how the division of surplus changes would be of interest. Unfortunately, comparisons to the static screening will generally be difficult due to the intractability of multidimensional static screening.

Third, how does screening across both space and time interact with competition between firms? \cite{ball2026competitive} considers multiple (single-good) firms that engage in sequential screening competing against one another. Their analysis follows a similar approach of considering the relaxed problem and finding equilibria before verifying global incentive constraints. Perhaps their results extend straightforwardly to a multi-good setting as the relaxed problem becomes separable across dimensions. In this world, several new questions can be considered. Under the optimal mechanism in this paper, upfront transfers were good-by-good but presented as aggregated. If there were multiple firms, consumers may want to only pay the upfront fee for some subset of goods at each of the many different firms. However, bundling upfront payments for strike prices across different goods may allow firms to lock in consumers. For example, buyers cannot just buy a membership to Costco's produce section; as a result of buying a Costco membership they purchase other goods in addition to the produce they came for. This also discourages the consumer from purchasing a membership at Costco's competitors (where once again, ``specialized'' memberships are not available). 

Fourth, many real-world sequential screening mechanisms feature a finite set of upfront and eventual strike prices, whereas the optimal mechanisms derived in this paper generally vary continuously and thus induce menus that have an uncountably infinite number of upfront and eventual strike prices. Characterizing optimal mechanisms when the menu size is restricted may further rationalize the pricing schemes seen across many industries.

Finally, many firms that interact with consumers over time do not use this type of mechanism. For example, as noted in \cite{yang_2025_nested}, streaming, tech, and AI firms often offer tiered subscriptions and zero spot prices. There are two commonalities between these firms: There are very low marginal costs to offering the service and interactions happen at a very high frequency. On the other hand, the examples considered in this paper feature larger marginal costs to providing the buyer with the good or service and fewer, but more impactful, interactions. I conjecture that incorporating these forces can provide a fuller description of sequential screening in the following ways. First, spot prices should be more sensitive to production costs compared to upfront payments. Second, if we consider replicating and re-scaling each good, upfront payments should remain unchanged while spot prices are scaled closer to zero. Both conjectures align with what is observed in practice and further exploration can enrich our understanding of optimal pricing schemes.

\newpage

\bibliography{cites}

\appendix

\newpage 

\section*{Appendix A: Omitted Proofs}
\makeatletter\def\@currentlabel{Appendix A}\makeatother
\label{Appendix A}

\subsubsection*{PROOF OF LEMMA \ref{formulation}}

\begin{proof}
    For any $u(\gamma, \theta)$, applying the envelope theorem to period two gives that the allocation rule must be $q(\gamma, \theta) = \nabla_\theta u(\gamma, \theta)$. As such, period two transfers are
    $$t_2(\gamma, \theta) = \theta \cdot \nabla_\theta u(\gamma, \theta) - u(\gamma, \theta).$$
    Let $U(\gamma'|\gamma)$ be the interim utility a consumer with type $\gamma$ receives when reporting $\gamma'$ and $U(\gamma) = U(\gamma|\gamma)$ be the interim utility of a type $\gamma$ consumer when reporting truthfully. Since $\gamma$ is not directly payoff-relevant and only affects payoffs by influencing the distribution of $\theta$, we have
    $$U(\gamma'|\gamma) = \int_\Theta u(\gamma',\theta) dF(\theta|\gamma) - t_1(\gamma'); U(\gamma) = \int_\Theta u(\gamma,\theta) dF(\theta|\gamma) - t_1(\gamma).$$
    Then, by incentive compatibility $\gamma \in \argmax_{\gamma'} U(\gamma'|\gamma)$ so the Envelope Theorem applies and local incentive constraints gives
    $$U_\gamma(\gamma) = U_\gamma(\gamma'=\gamma|\gamma) = \int_\Theta u(\gamma,\theta) f_\gamma(\theta|\gamma) d\theta.$$
    By the Fundamental Theorem of Calculus, we have
    $$U(\gamma) = U(\underline{\gamma}, \underline{\gamma}) + \int_{\underline{\gamma}}^\gamma U_\gamma(\gamma')d\gamma' = U(\underline{\gamma}, \underline{\gamma}) + \int_{\underline{\gamma}}^\gamma \int_\Theta u(\gamma',\theta) f_\gamma(\theta|\gamma') d\theta d\gamma'.$$
    As usual, we can set $U(\underline{\gamma}, \underline{\gamma}) = 0$. Equating this with the direct computation of $U(\gamma)$ and solving for transfers gives
    $$t_1(\gamma) = \int_\Theta u(\gamma,\theta) dF(\theta|\gamma) - \int_{\underline{\gamma}}^\gamma \int_\Theta u(\gamma',\theta) f_\gamma(\theta|\gamma') d\theta d\gamma'.$$
    Thus, $t_1, t_2, q$ are all pinned down by $u$ so the monopolist only needs to maximize over $u$; the monopolist's objective function can be written as 
    \begin{align*}
        \E[\gamma]{t_1(\gamma)} + \E[\gamma, \theta]{t_2(\gamma, \theta)} =& \int_\Gamma \left[\int_\Theta u(\gamma,\theta) dF(\theta|\gamma) - \int_{\underline{\gamma}}^\gamma \int_\Theta u(\gamma',\theta) f_\gamma(\theta|\gamma') d\theta d\gamma'\right] dG(\gamma) \\
        &+ \int_\Gamma \int_\Theta \left[\theta \cdot \nabla_\theta u(\gamma, \theta) - u(\gamma, \theta)\right] dF(\theta|\gamma) dG(\gamma).
    \end{align*}
    Observe that the first term of $\E[\gamma]{t_1(\gamma)}$ and the second term of $\E[\gamma, \theta]{t_2(\gamma, \theta)}$ cancel out: Any information rents given to the consumer for their post-contractual information can be extracted, in expectation, before that information arrives. The remainder of the computation is standard, interchanging the two integrals with respect to $\gamma$ to recover the standard hazard rate of $G$ terms: Expected transfers are
    \begin{align*}
        & \int_\Gamma\left[ -\int_{\underline{\gamma}}^\gamma \int_\Theta u(\gamma',\theta) f_\gamma(\theta|\gamma') d\theta d\gamma' + \int_\Theta \theta \cdot \nabla_\theta u(\gamma, \theta) f(\theta|\gamma) d\theta \right] g(\gamma) d\gamma \\
        =& \int_\Gamma \int_\Theta \Big[-u(\gamma, \theta) f_\gamma(\theta|\gamma) (1-G(\gamma)) + \theta \cdot \nabla_\theta u(\gamma, \theta) f(\theta|\gamma) g(\gamma) \Big] d\theta d\gamma \\
        =& \E[\gamma, \theta]{\theta \cdot \nabla_\theta u(\gamma, \theta) -u(\gamma, \theta) \frac{f_\gamma(\theta|\gamma)}{f(\theta|\gamma)} \frac{1-G(\gamma)}{g(\gamma)}}
    \end{align*}
    as desired. 
\end{proof}

\subsubsection*{PROOF OF LEMMA \ref{add_sep}}
\begin{proof}
    Suppose $F$ has invariant dependencies. Then, suppressing function arguments and using the fact that for any function $h(x), \frac{d \ln(h(x))}{dx} = h'(x)/h(x)$, at every $\theta$ and $\gamma$ we have
    \begin{align*}
        \nabla_\theta (\mathbf{V}f) &= \sum_j \left[f \frac{\partial \mathbf{V}^j}{\partial \theta^j} + \mathbf{V}^j \frac{\partial f}{\partial \theta^j}\right] = \sum_j \left[f \frac{\partial}{\partial \theta^j} \left(\frac{F_\gamma^j}{f^j}\right) + \frac{F_\gamma^j}{f^j} f \frac{\partial \ln(f)}{\partial \theta^j}\right] \\
        &= \sum_j \left[f \left(\frac{f^j f_\gamma^j - F_\gamma^j f^j_{\theta^j}}{(f^j)^2}\right) + \frac{F_\gamma^j}{f^j} f \frac{\partial}{\partial \theta^j} \left(\ln(c(F^1,...,F^n)) + \sum_\ell \ln(f^\ell)\right)\right] \\
        &= f \sum_j \left[ \left(\frac{f^j f_\gamma^j - F_\gamma^j f^j_{\theta^j}}{(f^j)^2}\right) + \frac{F_\gamma^j}{f^j} \frac{\partial \ln(c)}{\partial u^j} f^j + F_\gamma^j \frac{f^j_{\theta^j}}{(f^j)^2}\right] \\
        &= f\sum_j \left[ \frac{f_\gamma^j}{f^j} + F_\gamma^j \frac{\partial \ln(c)}{\partial u^j} \right] = f \frac{\partial}{\partial \gamma} \left[\sum_j \ln(f^j) + \ln(c(F^1,...,F^n))\right] \\
        &= f \frac{\partial \ln(f)}{\partial \gamma} = f \frac{f_\gamma}{f} = f_\gamma.
    \end{align*}
    Then for every $\gamma, \theta$, the product rule gives
    \begin{align*}
        \nabla_\theta \cdot \left(u(\gamma, \theta) \mathbf{V}(\theta|\gamma)f(\theta|\gamma)\right) &= \nabla_\theta u(\gamma, \theta) \cdot [\mathbf{V}(\theta|\gamma)f(\theta|\gamma)] + u(\gamma, \theta) [\nabla_\theta \cdot (\mathbf{V}(\theta|\gamma)f(\theta|\gamma))]
    \end{align*}
    so plugging the above into the following integral and applying the Divergence Theorem gives
    \begin{align*}
        \int_\Theta -u(\gamma, \theta)f_\gamma(\theta|\gamma)d\theta &= \int_\Theta - u(\gamma, \theta) [\nabla_\theta \cdot (\mathbf{V}(\theta|\gamma)f(\theta|\gamma))]d\theta \\
        &= \int_\Theta \nabla_\theta u(\gamma, \theta) \cdot [\mathbf{V}(\theta|\gamma)f(\theta|\gamma)] - \nabla_\theta \cdot \left(u(\gamma, \theta) \mathbf{V}(\theta|\gamma)f(\theta|\gamma)\right) d\theta \\
        &= \int_\Theta \nabla_\theta u(\gamma, \theta) \cdot [\mathbf{V}(\theta|\gamma)f(\theta|\gamma)] d\theta - \oint_{\partial \Theta} (u(\gamma, \theta) \mathbf{V}(\theta|\gamma)f(\theta|\gamma)) \cdot \mathbf{n} \: dS.
    \end{align*}
    Then, along the boundary $\partial \Theta$, for any $j$ we have that $F^j(\theta^j|\gamma)$ is either zero or one for all $\gamma$. As such, $F^j_\gamma(\theta^j|\gamma) = 0$ along the boundary and thus 
    $$(\mathbf{V}(\theta|\gamma)f(\theta|\gamma))^j = F^j_\gamma (\theta^j|\gamma) \prod_{i \neq j} f^i(\theta^i|\gamma) = 0$$
    so the boundary term vanishes for any $u$. Thus, we are left with 
    $$\int_\Theta -u(\gamma, \theta)f_\gamma(\theta|\gamma)d\theta = \int_\Theta \nabla_\theta u(\gamma, \theta) \cdot [\mathbf{V}(\theta|\gamma)f(\theta|\gamma)] d\theta;$$
    substituting this into the objective of Lemma \ref{formulation} and using $\nabla_\theta u(\gamma, \theta) = q(\gamma, \theta)$ gives that the principal seeks to maximize
    $$\E[\gamma, \theta]{\sum_j q^j(\gamma, \theta) \phi^j(\gamma, \theta^j)}.$$
\end{proof}

\subsubsection*{PROOF OF THEOREM \ref{no_bundlding}}
\begin{proof}
    Maximizing pointwise gives the optimal allocation rule to be
    $$q(\gamma, \theta) = \sum_{j=1}^n e^j I\left\{\phi^j(\gamma, \theta^j) \geq 0\right\}$$
    where $e^j$ is the $j$th basis vector and $I(A)$ is an indicator function over the set $A$. As virtual values are single-crossing in $\theta^j$ for each $j$ there is some threshold $\hat{\theta}^j(\gamma)$ such that 
    $$\phi^j(\gamma, \theta^j) \geq 0 \iff \theta^j \geq \hat{\theta}^j(\gamma).$$
    This is also the price $p^j(\gamma)$ a buyer reporting $\gamma$ faces when deciding whether or not to (eventually) purchase good $j$. Then, $q(\gamma, \theta)$ is cyclically monotone in $\theta$ for each $\gamma$ and thus induces a utility function $u(\gamma, \theta)$ that is 1-Lipschitz and convex in $\theta$. As virtual values are increasing in $\gamma$, the threshold $\hat{\theta^j}(\gamma)$ is decreasing in $\gamma$ so $q^j(\gamma, \theta)$ is increasing in $\gamma$. Using divergence theorem a la Lemma \ref{add_sep}, we can write
    $$U_\gamma(\gamma'|\gamma) = \int_\Theta u(\gamma, \theta)f_\gamma(\theta|\gamma) d\theta = \int_\Theta q(\gamma, \theta) \cdot [-F_\gamma(\theta|\gamma)] d\theta.$$
    As long as $q$ is increasing in $\gamma$, Assumption A1 gives that $-F_\gamma(\theta|\gamma) \geq 0$ so $U(\gamma'|\gamma)$ has increasing differences and thus global incentive constraints are satisfied. Next, integrating both sides and using separability of $q$ gives
    \begin{align*}
        U(\gamma) = \sum_j \int_{\underline{\gamma}}^\gamma \int_\Theta q^j(\gamma, \theta) [-F_\gamma^j(t|\gamma)] dt d\gamma = \sum_j \int_{\underline{\gamma}}^\gamma \int_{p^j(\gamma)}^{\bar{\theta}^j}-F_\gamma^j(t|\gamma) dt d\gamma
    \end{align*}
    while total expected gains from trade realized from a buyer with type $\gamma$ is 
    $$\int_\Theta u(\gamma, \theta) dF(\theta|\gamma) = \E[\theta|\gamma]{\sum_j \max\{\theta^j-p^j(\gamma), 0\}} = \sum_j \int_{p^j(\gamma)}^{\bar{\theta}^j}1-F^j(\theta^j|\gamma) d\theta^j.$$
    Period one transfers is equal to total surplus minus consumer surplus, so 
    \begin{align*}
        t_1(\gamma) &= \sum_j \int_{p^j(\gamma)}^{\bar{\theta}^j}1-F^j(\theta^j|\gamma) d\theta^j - \left[\sum_j \int_{\underline{\gamma}}^\gamma \int_{p^j(\gamma)}^{\bar{\theta}^j}-F_\gamma^j(t|\gamma) dt d\gamma\right] \\
        &= \sum_{j=1}^n \left[\int_{p^j(\gamma)}^{\bar{\theta}^j}1-F^j(\theta^j|\gamma) d\theta^j + \int_{\underline{\gamma}}^\gamma \int_{p^j(\gamma)}^{\bar{\theta}^j}F_\gamma^j(t|\gamma') dt d\gamma' \right].
    \end{align*}
\end{proof}

\subsubsection*{PROOF OF PROPOSITION \ref{ironing}}

\begin{proof}
    The proof proceeds in two steps. First, we show that under independence, it is without loss to consider allocation rules that are monotone in each dimension. Second, higher values of $\gamma$ correspond to pointwise higher values of $q$ when solving the problem $\gamma$-by-$\gamma$.

    \begin{lemma}\label{ironing_lem}
        For every cyclically monotone $q(\gamma, \theta)$, there exists $\{\hat{q}^j(\gamma, \theta^j)\}$ such that $q$ and $\{\hat{q}^j(\gamma, \theta^j)\}$ produce the same surplus. Conversely, every collection $\{\hat{q}^j\}$ produces a cyclically monotone $q$ when concatenated.
    \end{lemma}

    \begin{proof}
        Suppose $q$ is cyclically monotone. We have that firm profits are 
        \begin{align*}
            \E[\gamma, \theta]{\sum_j q^j(\gamma, \theta) \phi^j(\gamma, \theta^j)} &= \sum_j \E[\gamma, \theta]{ q^j(\gamma, \theta) \phi^j(\gamma, \theta^j)} \\
            &= \sum_j \E[\gamma, \theta^j]{\E[\theta^{-j}|\gamma]{ q^j(\gamma, \theta)}\phi^j(\gamma, \theta^j)}
        \end{align*}
        so let 
        $$\hat{q}^j(\gamma, \theta^j) = \E[\theta^{-j}|\gamma]{ q^j(\gamma, \theta)} = \int_{\Theta^{-j}} q^j(\gamma, \theta) \prod_{k \neq j} f^k(\theta^k|\gamma)d\theta^{-j}.$$
        Let $\hat{\theta}^j > \theta^j$; we have that 
        $$\hat{q}^j(\gamma, \hat{\theta}^j) - \hat{q}^j(\gamma, \theta^j) =  \int_{\Theta^{-j}} \left(q^j(\gamma, \hat{\theta}^j, \theta^{-j})-q^j(\gamma, \theta^j, \theta^{-j})\right) \prod_{k \neq j} f^k(\theta^k|\gamma)d\theta^{-j}.$$
        Applying cyclic monotonicity of $q$ to the two-cycle $(\hat{\theta}^j, \theta^{-j}); (\theta^j, \theta^{-j})$ for arbitrary $\theta^{-j}$ gives that 
        $$q^j(\gamma, \hat{\theta}^j, \theta^{-j}) \geq q^j(\gamma, \theta^j, \theta^{-j}) \implies q^j(\gamma, \hat{\theta}^j, \theta^{-j})-q^j(\gamma, \theta^j, \theta^{-j}) \geq 0$$
        for all $\theta^{-j}$. Thus, integrating over $\Theta^{-j}$ gives that $\hat{q}^j(\gamma, \theta^j)$ is monotone in $\theta^j$. By construction, $\{\hat{q}^j(\gamma, \theta^j)\}$ produces the same surplus as $q$. The second statement of the Lemma is standard.
    \end{proof}
    
    By Lemma \ref{ironing_lem}, it suffice to optimize over $\{q^j(\gamma, \theta^j)\}$ subject to (standard) monotonicity in $\theta^j$ opposed to cyclically monotone $q$. Let 
    $$W^j(\gamma, \theta^j) = \theta^j f^j(\theta^j|\gamma) g(\gamma) + F^j_\gamma(\theta^j|\gamma)(1-G(\gamma))$$
    so the regularity in $\gamma$ assumption requires $W^j$ to be increasing in $\gamma$. The firm's objective can be written as 
    \begin{align*}
        \max_{\{q^j\}} & \sum_j \int_\Gamma \int_{\Theta^j} q^j(\gamma, \theta^j) W^j(\gamma, \theta^j) d\theta^j d\gamma \\
        \text{s.t. } & \: q^j(\gamma, \theta^j) \text{ increasing in } \theta^j; \\
        & \text{ IC1}.
    \end{align*}
    Note here that analytically solving the problem in each dimension would require ironing. However, the proof of the proposition will not do any ironing; we will directly characterize the solution using monotone comparative statics. 

    Fix $j$. Monotone functions $q^j(\gamma, \cdot): \Theta^j \to [0, 1]$ form a lattice under pointwise maximum/minimum. The set $\Gamma$ is straightforwardly a lattice. It suffice to show that 
    $$O(q, \gamma) = \int_{\Theta^j} q(\theta^j) W^j(\gamma, \theta^j) d\theta^j$$
    is supermodular in $q, \gamma$. Fix $q, q'; \gamma, \gamma'$. Without loss of generality, suppose $\gamma \geq \gamma'$ since $\Gamma$ is one-dimensional. Note that for any $a, b, \max\{a, b\} = a+b-\min\{a, b\}$. We have
    \begin{align*}
        & O(\max\{q, q'\}, \max\{\gamma, \gamma'\}) + O(\min\{q, q'\}, \min\{\gamma, \gamma'\}) \\
        = & \int_{\Theta^j} \max\{q(\theta^j), q'(\theta^j)\} W^j(\gamma, \theta^j)d\theta^j + \int_{\Theta^j}\min\{q(\theta^j), q'(\theta^j)\} W^j(\gamma', \theta^j) d\theta^j \\
        = & \int_{\Theta^j} \left[q(\theta^j) + q'(\theta^j) - \min\{q(\theta^j), q'(\theta^j)\}\right]W^j(\gamma, \theta^j)d\theta^j + \int_{\Theta^j}\min\{q(\theta^j), q'(\theta^j)\} W^j(\gamma', \theta^j) d\theta^j \\ 
        = &\int_{\Theta^j} q(\theta^j) W(\gamma, \theta^j)d\theta^j + \int_{\Theta^j} \left[q'(\theta^j) - \min\{q(\theta^j), q'(\theta^j)\}\right] W^j(\gamma, \theta^j)d\theta^j \\
        & + \int_{\Theta^j}\min\{q(\theta^j), q'(\theta^j)\} W^j(\gamma', \theta^j) d\theta^j \\
        \geq &\int_{\Theta^j} q(\theta^j) W(\gamma, \theta^j)d\theta^j + \int_{\Theta^j} \left[q'(\theta^j) - \min\{q(\theta^j), q'(\theta^j)\}\right] W^j(\gamma', \theta^j)d\theta^j \\
        &+ \int_{\Theta^j}\min\{q(\theta^j), q'(\theta^j)\} W^j(\gamma', \theta^j) d\theta^j \\
        = &\int_{\Theta^j} q(\theta^j) W(\gamma, \theta^j)d\theta^j + \int_{\Theta^j} q'(\theta^j) W^j(\gamma', \theta^j)d\theta^j \\
        = & O(q, \gamma) + O(q', \gamma')
    \end{align*}
    as desired. The inequity comes from the fact that pointwise $\left[q'(\theta^j) - \min\{q(\theta^j), q'(\theta^j)\}\right] \geq 0$ so $W$ being increasing in $\gamma$ gives that replacing $W^j(\gamma, \theta^j)$ with $W^j(\gamma', \theta^j)$ makes the value smaller.
\end{proof}

\subsubsection*{PROOF OF PROPOSITION \ref{seq_irrelevance}}

\begin{proof}
    Let $t(\gamma, \theta)$ be net transfers. First, any $q(\gamma, \theta), t(\gamma, \theta)$ that is implementable in the original setting is implementable in the sequential setting.\footnote{Suppose that goods are allocated at the end but information must be reported as soon as it arrives, so the allocation of good $j$ could, in theory, depend on the buyer's reported value of $\theta^k$ even if $k > j$.} We will prove the contrapositive: If $q(\gamma, \theta), t(\gamma, \theta)$ is not implementable in the sequential setting, then it must not be implementable in the original setting. 

    Suppose $q(\gamma, \theta), t(\gamma, \theta)$ is not implementable in the sequential setting. Then, the buyer must have a profitable deviation at some point along the path of play of the mechanism. For notational simplicity, suppose the deviation happens after $\gamma$ has been truthfully reported; the same reasoning holds for deviations in $\gamma$ as well. Thus, there exists some $\gamma, \theta^1,...,\theta^k$ such that the buyer has learned $\gamma, \theta^1,...,\theta^k$, truthfully reported $\gamma, \theta^1,...,\theta^{k-1}$, and can profitably deviate to a sequence of misreports $\{\tilde{\theta}^j(\theta^1,...,\theta^j)\}_{j=k}^n$. Slightly abusing notation, let $\tilde{\theta}(\theta)$ be the buyer's reports for a given realization of $\theta$. For this to be profitable, we have
    $$\E[\theta^{k+1},...,\theta^n|\gamma]{\theta \cdot q(\gamma, \tilde{\theta}(\theta))- t(\gamma, \tilde{\theta}(\theta))\bigg|\theta^1,..,\theta^k} > \E[\theta^{k+1},...,\theta^n|\gamma]{\theta \cdot q(\gamma, \theta)- t(\gamma, \theta)\bigg|\theta^1,..,\theta^k}.$$
    A necessary condition for the above inequality to hold is for there to exist a realization $\theta$ for which 
    $$\theta \cdot q(\gamma, \tilde{\theta}(\theta))- t(\gamma, \tilde{\theta}(\theta)) > \theta \cdot q(\gamma, \theta)- t(\gamma, \theta).$$
    However, this implies that a buyer with realized values $\gamma, \theta$ as above has a profitable deviation to report $\gamma, \tilde{\theta}(\theta)$, as desired.

    Then, the monopolist's objective in the sequential case, as formalized in \cite{segal2014}, coincides with the monopolist's objective in the original setting, albeit with different constraints. Since pointwise maximization is implementable in the original problem, it is also implementable in the sequential case. 
\end{proof}

\subsubsection*{PROOF OF PROPOSITION \ref{same_rents}}

\begin{proof}
    Suppose $q, t_1, t_2$ is incentive compatible in the original problem and let $\hat{q}(\gamma, z) = q(\gamma, v(\gamma, z)), \hat{t}(\gamma) = \E[z]{t_1(\gamma) + t_2(\gamma, v(\gamma, z))}$. Let 
    $$\hat U(\gamma'|\gamma) = \E[z]{\hat{q}(\gamma', z) \cdot v(\gamma, z)} - \hat{t}(\gamma'), \hat U(\gamma) = \hat U(\gamma|\gamma)$$
    be the buyer's interim utility function from reporting $\gamma'$ at $\gamma$ and their interim utility when truthful. By the Envelope Theorem and interchanging differentiation and expectation, 
    $$\hat U_\gamma(\gamma) = \E[z]{\hat{q}(\gamma, z) \cdot v_\gamma(\gamma, z)}.$$
    Then, applying the Fundamental Theorem of Calculus gives
    $$\E[z]{\hat{q}(\gamma, z) \cdot v(\gamma, z)} - \hat{t}(\gamma) = \hat{U}(\gamma) = \hat U(\underline{\gamma}) + \int_{\underline{\gamma}} \hat{U}_\gamma(\gamma')d\gamma'$$
    so transfers are now pinned down by (setting $\hat U(\underline{\gamma}) = 0$):
    \begin{align*}
        \hat{t}(\gamma) &= \E[z]{\hat{q}(\gamma, z) \cdot v(\gamma, z) - \int_{\underline{\gamma}}^\gamma \hat{q}(\gamma', z) \cdot v_\gamma(\gamma', z) d\gamma'}.
    \end{align*}
    Using the standard technique of interchanging the two integrals taken with respect to $\gamma$, expected transfers can be written as
    $$\E[\gamma]{\hat{t}(\gamma)} = \E[\gamma, z]{\hat{q}(\gamma, z) \cdot \left(v(\gamma, z) - v_\gamma(\gamma, z)\frac{1-G(\gamma)}{g(\gamma)}\right)}.$$
    Then, $z = v^{-1}(\gamma, \theta)$ so taking $q(\gamma, \theta) = \hat{q}(\gamma, v^{-1}(\gamma, \theta))$ and $\mathbf{V}(\theta|\gamma) = v_\gamma(\gamma, v^{-1}(\gamma, \theta))$ gives expected transfers to be
    $$\E[\gamma]{\hat{t}(\gamma)} = \E[\gamma, \theta]{q(\gamma, \theta) \cdot \left(\theta - \mathbf{V}(\theta|\gamma) \frac{1-G(\gamma)}{g(\gamma)}\right)}.$$
    As $q(\gamma, \theta)$ were implementable in the original problem, $q(\gamma, \theta) = \nabla_\theta u(\gamma, \theta).$ Then, expected profits as formulated above coincides with Equation \ref{eqn1} so it suffice to show that $\mathbf{V}(\theta|\gamma)$ satisfies $-\nabla_\theta \cdot (\mathbf{V}(\theta|\gamma) f(\theta|\gamma)) = f_\gamma(\theta|\gamma)$ and boundary conditions vanish.

    By construction, $v$ satisfies
    $$F(v(\gamma, z)|\gamma) = \prod_j z^j$$
    so differentiating both sides with respect to $\gamma$ and replacing $z$ with $\theta$ gives
    $$\nabla_\theta F(v(\gamma, z)|\gamma) \cdot v_\gamma(\gamma, \theta) + F_\gamma(v(\gamma, z)|\gamma) = 0 \implies F_\gamma(\theta|\gamma) = - \sum_{i=1}^n \frac{\partial F}{\partial \theta^i} \mathbf{V}^i(\gamma, \theta).$$
    Using this identity, we have
    \begin{align*}
        f_\gamma(\theta|\gamma) &= \frac{\partial^n}{\partial \theta^1,...,\partial \theta^n}\left[F_\gamma(\theta|\gamma)\right] = \frac{\partial^n}{\partial \theta^1,...,\partial \theta^n}\left[- \sum_{i=1}^n \frac{\partial F}{\partial \theta^i} \mathbf{V}^i(\gamma, \theta)\right] = -\nabla_\theta \cdot (\mathbf{V}(\theta|\gamma) f(\theta|\gamma)).
    \end{align*}
    Finally, boundary conditions vanish as for $\theta \in \partial \Theta$, we have that $v^{-1}(\gamma,\theta) \in \partial [0, 1]^n$ and if each entry of $z$ is either zero or one then
    $$v^j(\gamma, z) = \begin{cases}
        \underline{\theta}^j & \text{ if } z^j = 0;\\
        \bar{\theta}^j & \text{ if } z^j = 1.
    \end{cases}$$
    As such, $v(\gamma, z)$ is constant in $\gamma$ so $v_\gamma(\gamma, z) = 0$ on $\partial [0, 1]^n$ and boundary terms vanish.
\end{proof}

\subsubsection*{PROOF OF PROPOSITION \ref{yes_orthog}}

\begin{proof}
    Suppose $F$ satisfies invariant dependencies. Then, we can write
    $$\theta = F^{-1}(C(F^1(\theta^1|\gamma),..., F^n(\theta^n|\gamma))|\gamma) = F^{-1}(\eta|\gamma)$$
    for $\eta \sim C$. We also have $\theta = v(\gamma, z)$. Let $\kappa$ be a copula so $\eta = \kappa(z)$. Then, 
    $$v^j(\gamma, z) = (F^j)^{-1}(\kappa^j(z)|\gamma)$$
    and
    $$v^{-1}(\gamma, \theta) = z = \kappa^{-1}((F^1)^{-1}(\theta^1|\gamma), ..., (F^n)^{-1}(\theta^n|\gamma))$$
    which also implies 
    $$\kappa^j(v^{-1}(\gamma, \theta)) = F^j(\theta^j|\gamma).$$
    Then,
    \begin{align*}
        \mathbf{V}^j(\gamma, \theta) &= v_\gamma^j(\gamma, v^{-1}(\gamma, \theta)) = \frac{\partial}{\partial \gamma}\left[v^j(\gamma, z)\right]_{z = v^{-1}(\gamma, \theta)} \\
        &= \frac{\partial}{\partial \gamma}\left[(F^j)^{-1}(\kappa^j(z)|\gamma)\right]_{z = v^{-1}(\gamma, \theta)} \\
        &= \frac{\partial}{\partial \gamma}\left[(F^j)^{-1}(\eta^j|\gamma)\right]_{\eta^j = \kappa^j(v^{-1}(\gamma, \theta))} \\
        &= \frac{\partial}{\partial \gamma}\left[(F^j)^{-1}(\eta^j|\gamma)\right]_{\eta^j = F^j(\theta^j|\gamma)} \\
        &= -\frac{F^j_\gamma(\theta^j|\gamma)}{f^j(\theta^j|\gamma)}
    \end{align*}
    where the last line comes from the inverse function theorem. Thus, the objective found via Proposition \ref{same_rents} is equivalent to the objective found via Theorem \ref{no_bundlding} and the relaxed problem thus has the same solution as the original problem. Thus, the pointwise optimal allocation rule solves both problems.
\end{proof}

\end{document}